\documentclass[aps,prd,twocolumn,preprintnumbers,amsmath,amssymb,
superscriptaddress,showpacs,floatfix]{revtex4}

\usepackage{graphicx}
\usepackage{dcolumn}
\usepackage{bm}
\usepackage{epsfig}
\usepackage{graphics}
\usepackage{latexsym}
\usepackage{subfigure}

\def\beq{\begin{equation}}
\def\eeq{\end{equation}}
\def\bea{\begin{eqnarray}}
\def\eea{\end{eqnarray}}
\newcommand{\beqs}{\begin{subequations}}
\newcommand{\eeqs}{\end{subequations}}

\newcommand\eq[1]{Eq.~(\ref{#1})}
\newcommand\eqs[2]{Eqs.~(\ref{#1}) and (\ref{#2})}

\def\lf{\left(}
\def\rg{\right)}

\newcommand{\Eref}[1]{Eq.~(\ref{#1})}
\newcommand{\Sref}[1]{Sec.~\ref{#1}}
\newcommand{\Fref}[1]{Fig.~\ref{#1}}
\newcommand{\Tref}[1]{Table~\ref{#1}}
\newcommand{\cref}[1]{Ref.~\cite{#1}}

\renewcommand{\Re}{\mathsf{Re}}
\renewcommand{\Im}{\mathsf{Im}}
\newcommand{\hh}{{\ensuremath{I{\kern-2.6pt h}}}}
\newcommand{\bhh}{{\ensuremath{\bar{I{\kern-2.6pt h}}}}}

\newcommand{\gev}{~\text{GeV}}
\newcommand{\tev}{~\text{TeV}}
\newcommand{\GeV}{\text{GeV}}
\newcommand{\TeV}{\text{TeV}}

\newcommand{\omg}{\Omega_{\mathrm{LSP}}h^2}
\newcommand{\cdm}{\Omega_{\mathrm{CDM}}h^2}
\newcommand{\damu}{\delta a_\mu}
\newcommand{\dstau}{\Delta_{\tilde{\tau}_1}}
\newcommand{\dch}{\Delta_{\tilde{\chi}^\pm}}
\newcommand{\dnt}{\Delta_{\tilde{\chi}_2}}
\newcommand{\ssi}{\sigma_{\tilde{\chi}p}^{\mathrm{SI}}}
\newcommand{\xssi}{\xi\sigma_{\tilde{\chi}p}^{\mathrm{SI}}}
\newcommand{\ssd}{\sigma_{\tilde{\chi}p}^{\mathrm{SD}}}
\newcommand{\amg}{A_0/M_{1/2}}

\newcommand{\sufour}{SU(4)_\mathrm{c}}
\newcommand{\sutwo}[1]{SU(2)_\mathrm{#1}}
\newcommand{\DR}{$\overline{\mathrm{DR}}~$}
\newcommand{\MS}{$\overline{\mathrm{MS}}~$}
\newcommand{\neutralino}{\tilde{\chi}}

\newcommand{\staua}{\tilde{\tau}_1}
\newcommand{\sutop}{\tilde{t}}

\newcommand{\bsmumu}{B_s\rightarrow\mu^+\mu^-}
\newcommand{\bsmm}{{\ensuremath{{\rm BR}\lf B_s\to \mu^+\mu^-\rg}}}
\newcommand{\bsg}{{\ensuremath{{\rm BR}\lf b\to s\gamma\rg}}}
\newcommand{\btn}{{\ensuremath{{\rm R}\lf B_u\to \tau\nu\rg}}}

\newcommand{\mx}{{\ensuremath{m_{\rm LSP}}}}
\newcommand{\mgl}{{\ensuremath{m_{\tilde g}}}}
\newcommand{\mch}{{\ensuremath{m_{\tilde \chi^\pm}}}}
\newcommand{\ch}{{\ensuremath{{\tilde \chi^\pm}}}}
\newcommand{\cha}{{\ensuremath{{\tilde \chi^+_1}}}}
\newcommand{\mh}{{\ensuremath{m_{h}}}}
\newcommand{\mo}{{\ensuremath{m_{0}}}}

\newcommand{\dew}{{\ensuremath{\Delta_{\rm EW}}}}
\newcommand{\mst}{{\ensuremath{m_{\tilde\tau_1}}}}
\newcommand{\mg}{{\ensuremath{M_{1/2}}}}
\newcommand{\xx}{{\ensuremath{\tilde\chi}}}
\newcommand{\xxb}{{\ensuremath{\tilde\chi_2}}}
\newcommand{\xstau}{{\ensuremath{\tilde\chi-\tilde\tau_1}}}
\newcommand{\tnb}{{\ensuremath{\tan\beta}}}
\newcommand{\sign}{{\ensuremath{\rm sign}}}

\begin{document}

\preprint{UT-STPD-15/01}

\title{Probing the hyperbolic branch/focus point region of the
constrained minimal supersymmetric standard model with generalized
Yukawa quasi-unification}

\author{N. Karagiannakis}
\email{nikar@auth.gr} \affiliation{School of Electrical and
Computer Engineering, Faculty of Engineering, Aristotle University
of Thessaloniki, Thessaloniki 54124, Greece}
\author{G. Lazarides}
\email{lazaride@eng.auth.gr}
\affiliation{School of Electrical and Computer
Engineering, Faculty of Engineering, Aristotle University of Thessaloniki,
Thessaloniki 54124, Greece}

\author{C. Pallis}
\email{cpallis@ific.uv.es} \affiliation{Departament de F\'{i}sica
Te\`{o}rica and IFIC, Departament de F\'{i}sica Te\`{o}rica and
IFIC, E-46100 Burjassot, SPAIN}

\date{\today}

\begin{abstract}
We analyze the parametric space of the constrained minimal
supersymmetric standard model with $\mu>0$ supplemented by a
generalized asymptotic Yukawa coupling quasi-unification condition
which yields acceptable masses for the fermions of the third
family. We impose constraints from the cold dark matter abundance
in the universe and its direct detection experiments, the
$B$-physics, as well as the masses of the sparticles and the
lightest neutral CP-even Higgs boson. Fixing the mass of the
latter to its central value from the LHC and taking
$40\lesssim\tnb\lesssim50$, we find a relatively wide allowed
parameter space with $-11\lesssim\amg\lesssim15$ and mass of the
lightest sparticle in the range $(0.09-1.1)~\TeV$. This sparticle
is possibly detectable by the present cold dark matter direct
search experiments. The required fine-tuning for the electroweak
symmetry breaking is much milder than the one needed in the
neutralino-stau coannihilation region of the same model.
\end{abstract}

\pacs{\scriptsize 12.10.Kt, 12.60.Jv, 95.35.+d, 14.80.Cp; \hfill
{\sl\bfseries Published in} {Phys. Rev. D} {\bf 92}, no. 8, 085018
(2015)}
%\keywords{MSSM, Hyperbolic Branch, Focus Point, Higgs Boson, Dark
%Matter, Neutralino, Higgsino}
%Use showkeys class option if keyword display desired
\maketitle

\section{Introduction}

The well-known \textit{constrained minimal supersymmetric standard
model} (CMSSM)
\cite{chamseddine1982,nath1983,ohta1983,hall1983,arnowitt1992,ross1992,barger1994,kane1994}
is a highly predictive version of the \textit{minimal
supersymmetric standard model} (MSSM). Its basic characteristic is
that it employs universal boundary conditions for the soft
\textit{supersymmetry} (SUSY) breaking terms. Namely, the free
parameters of the CMSSM are the following
\begin{equation}
\sign\mu,~~\tan\beta,~~\mg,~~m_0,~~\mbox{and}~~A_0, \label{par}
\end{equation}
where $\sign\mu$ is the sign of $\mu$, the mass parameter mixing
the electroweak Higgs superfields $H_2$ and $H_1$ of the MSSM
which couple to the up- and down-type quarks respectively, $\tnb$
is the ratio of the vacuum expectation values
of $H_2$ and $H_1$, while the remaining symbols denote the
common gaugino mass, the common scalar mass, and the common
trilinear scalar coupling constant, respectively, defined at the
\emph{grand unified theory} (GUT) scale $M_{\rm GUT}$,
which is determined by the unification of the gauge coupling
constants.

However, the recently announced experimental data on the mass of
the \emph{standard model} (SM)-like Higgs boson
\cite{Aad:2014aba,CMS:2014ega} and the branching ratio $\bsmm$
of the process $B_s\to\mu^+\mu^-$
\cite{LHCb2012,albrecht2012,LHCb2013} in conjunction with the
upper bound \cite{Ade:2013zuv} from \emph{cold dark matter}
(CDM) considerations on the relic abundance $\omg$ of the
\emph{lightest SUSY particle} (LSP) -- which is the
lightest neutralino $\xx$ -- put under considerable stress
\cite{Ellis2012,baer2012,Buchmueller:2013rsa,ros15} the parameter
space of the CMSSM. It is by now clear, however, that there
still exist viable slices of the following three regions of the
parameter space of the CMSSM which give acceptable values of
$\omg$:

\begin{enumerate}

\item[(i)] \textit{The neutralino-stau ($\xstau$) coannihilation
region} with $\mg\gg m_0$, where $\xx$ is an almost pure bino and
$\omg$ is reduced to an acceptable level due to the proximity
between the mass $\mx$ of the LSP and the mass $\mst$ of the
lightest stau $\staua$, which is the \textit{next-to-LSP} (NLSP).

\item[(ii)] \textit{The A-funnel region} \cite{Ellis2012}
appearing at large ($\gtrsim40$) $\tnb$ values, where the mass
$m_A$ of the CP-odd Higgs boson $A$ satisfies the relation
$m_A\simeq 2\mx$ and so the $\xx-\xx$ pair annihilation procedure
is enhanced by an $A$-pole exchange in the $s$-channel, thereby
reducing $\omg$.

\item[(iii)] \textit{The hyperbolic branch/focus point (HB/FP)
region} \cite{chan1998, baerfocus, nathfocus,feng, feng12,
focus2014, profumo, akula2012,nath13} located at large values of
$m_0\gg\mg$ and small $\mu$'s, which ensures a sizable higgsino
fraction of $\xx$. The $\omg$ remains under control for
$\mx\lesssim1~\TeV$ thanks to the rapid $\xx-\xx$ annihilation and
the neutralino-chargino ($\xx/\xxb-\cha$) coannihilation
\cite{baerCA,darksusy,microfp} ($\xxb$ is the next-to-lightest
neutralino and $\cha$ the lightest chargino).

\end{enumerate}

It would be interesting to investigate whether this available
parameter space survives in even more restrictive versions of the
CMSSM, which can emerge by embedding it in concrete SUSY GUT
models. Here we focus on the \textit{Pati-Salam} (PS) GUT model
based on the gauge group $G_\mathrm{PS}=
\sufour\times\sutwo{L}\times\sutwo{R}$ \cite{antoniadis1989,
jeannerot2000}. It is interesting to note that this model can
arise \cite{leontaris,leontaris1} from the standard weakly coupled
heterotic string and the D-brane setups \cite{brane,brane1} and
its phenomenology has been extensively studied -- see e.g.
Ref.~\cite{phen}. It can also naturally arise in the framework of
non-commutative geometry \cite{noncom}.

The SUSY PS model in its simplest realization leads
\cite{lazarides1994} to `asymptotic' \textit{Yukawa unification}
(YU) \cite{ananthanarayan1991,Gomez:1999dk}, i.e. the exact
unification, at the GUT scale $M_\mathrm{GUT}$, of the Yukawa
coupling constants $h_t$, $h_b$, and $h_\tau$ of the top quark
$t$, the bottom quark $b$, and the tau lepton $\tau$,
respectively. Although the CMSSM with YU is an elegant model, it
yields unacceptable values of the $b$-quark mass $m_b$ for both
signs of $\mu$ given the experimental values of the $t$ and $\tau$
masses -- which, combined with YU, naturally restrict $\tan\beta$
to large values. This is due to the presence of sizable SUSY
corrections
\cite{carena1994,hempfling1994,hall1994,anandakrishnan2014} to the
tree-level $b$-quark mass (about 20\%), which drive it well above
[a little below] its 95\% \textit{confidence level} (c.l.)
experimental range for $\mu>0$ $[\mu<0]$ -- we use the standard
convention of \cref{skands2004} for the sign of $\mu$.

In order to circumvent this difficulty, we consider here the
extended PS SUSY GUT model introduced in Ref.~\cite{gomez2002} and
reviewed in Refs.~\cite{Lazarides:2004pq,karagiannakis2013}, which
yields a moderate
deviation from exact YU and, thus, allows acceptable values of
the $b$-quark mass for both signs of $\mu$ within the CMSSM. In
particular, the Higgs sector of the simplest PS model
\cite{antoniadis1989,jeannerot2000} is extended so that $H_2$ and
$H_1$ are not exclusively contained in a $SU(4)_c$ singlet,
$SU(2)_{\rm L}\times SU(2)_{\rm R}$ bidoublet superfield, but receive
subdominant contributions from another bidoublet too which belongs
to the adjoint representation of $SU(4)_c$. As a result, YU is
naturally violated and replaced by a set of asymptotic Yukawa
quasi-unification (YQU) conditions:
\begin{eqnarray}\label{yquc}
 h_t(M_\mathrm{GUT}):h_b(M_\mathrm{GUT}):h_\tau(M_\mathrm{GUT})=\nonumber\\
\left|\frac{1-\rho\alpha_2/\sqrt{3}}{\sqrt{1+|\alpha_2|^2}}\right|:
\left|\frac{1-\rho\alpha_1/\sqrt{3}}{\sqrt{1+|\alpha_1|^2}}\right|:
\left|\frac{1+\sqrt{3}\rho\alpha_1}{\sqrt{1+|\alpha_1|^2}}\right|.
\end{eqnarray}
These conditions depend on two complex parameters $\alpha_1$,
$\alpha_2$ and one real and positive parameter $\rho$. The
parameters $\alpha_1$ and $\alpha_2$ describe the mixing of the
components of the $SU(4)_\mathrm{c}$ singlet and 15-plet Higgs
bidoublets, while $\rho$ is the ratio of their respective Yukawa
coupling constants to the fermions of the third family
\cite{karagiannakis2012,karagiannakis2013}.

 In our original papers \cite{gomez2002,gomez2003,
karagiannakis2011,karagiannakis2011b,karagiannakis2012b}, we have
considered monoparametric versions of the YQU conditions depending
on just one parameter, which was considered for simplicity real
and replaced $\tnb$ in the investigation of the parametric space
of the model. These forms of the YQU conditions arise by taking
$\alpha_1=-\alpha_2$ for $\mu>0$ or $\alpha_1=\alpha_2$ for
$\mu<0$. Indeed, these choices turn out to be suitable for
generating an adequate violation of YU and ensuring, at the same
time, successful radiative \textit{electroweak symmetry breaking}
(EWSB) and the presence of a neutralino LSP in a large fraction of
the parametric space. However, the emergent versions of CMSSM are
by now experimentally excluded. For $\mu<0$, this is due to the
incompatibility between the bound on the branching ratio $\bsg$ of
the process $b\to s\gamma$ and the CDM constraints
\cite{gomez2003}. For $\mu>0$, on the other hand, it is due to the
fact that the parameter space where the CDM constraint on $\omg$
is satisfied thanks to $\xstau$ coannihilations turns out to be
non-overlapping with the space allowed by the data
\cite{LHCb2012,albrecht2012} on $\bsmm$ \cite{karagiannakis2012b}.
The main reason for the latter negative result is that
$\tan\beta$ remains large and, hence, it enhances the SUSY
contribution to $\bsmm$.

In order to overcome this hurdle, we adopted in
Refs.~\cite{karagiannakis2012,karagiannakis2013} the more general
version of \Eref{yquc} without imposing any relation between
$\alpha_1$ and $\alpha_2$. As a consequence, we could accommodate
more general values of the ratios $h_m/h_n$ with $m,n=t,b,\tau$ --
still of order unity for natural values of the model parameters --
and lower $\tnb$'s. The allowed parameter space of the model found
in \cref{karagiannakis2012} is then mainly determined by the
interplay between the constraints from the $\bsmm$, CDM, and the
results of LHC on the Higgs boson $h$ mass $m_h$. Actually, our
acceptable solutions are due to a synergy between $\xstau$
coannihilation and the $A$-funnel mechanism -- see points (i) and
(ii) above -- given that $m_H\simeq m_A$ ($H$ is the heavier
CP-even neutral Higgs boson with mass $m_H$). However, $\mx$ comes
out to be large ($\sim1~\TeV$), which makes the direct
detectability of the LSP very difficult and the sparticle spectrum
very heavy. Furthermore, the emergent values of $\mg,~\mo$, and
$\mu$ lie in the multi-TeV range, which puts under some stress the
naturalness of the radiative EWSB.

Therefore, the tantalizing question, which we address here, is
whether the conditions in \Eref{yquc} can become compatible with
the solutions obtained in the HB/FP region of the CMSSM -- see
point (iii) above. We show that this is not only possible but also
quite promising since, apart from allowing a much wider parametric
space, it gives smaller values of $\mu$ and $\mx$, despite the
fact that $\mo$ and $\mg$ are quite large and the sfermions,
sleptons and heavier Higgs bosons are quite heavy. Moreover, the
LSP can be probed by the CDM detection experiments, which play a
crucial role in limiting the parameter space of the model.
Finally, the deviation needed from exact YU can be naturally
attributed to the YQU conditions in \Eref{yquc}, while the
fine-tuning required for the radiative EWSB turns out to be milder
than the one needed in the $\xstau$ coannihilation region for
$\mx$'s yielding $\omg$'s close to the cosmological upper bound.

We first (Sec.~\ref{sec:pheno}) review the phenomenological
and cosmological requirements which we consider in our
investigation. Next (\Sref{hbfp}), we exhibit the salient features
of the HB/FP region of the CMSSM and find, in Sec.~\ref{res}, the
resulting restrictions on the parameters of the CMSSM.  We,
finally, check, in \Sref{yuk}, the consistency with \Eref{yquc}
and discuss the naturalness of the model in \Sref{nat}. We
summarize our conclusions in Sec.~\ref{con}.

\section{Phenomenological and Cosmological  Constraints}
\label{sec:pheno}

In our investigation, we closely follow the renormalization group
and radiative EWSB analysis of
Refs.~\cite{karagiannakis2012,karagiannakis2013}. Most notably, we
use an optimal, common supersymmetric threshold
\begin{equation}
 \label{opt} M_\mathrm{SUSY}\simeq(m_{\tilde{t}_1}
m_{\tilde{t}_2})^{1/2}
\end{equation}
(with $\tilde{t}_{1,2}$ being the stop quark mass eigenstates)
which reduces drastically the one-loop corrections to the Higgs
boson masses, making their calculation from the tree-level
effective potential quite accurate and stable \cite{barger1994}.
The SUSY spectrum is also evaluated at $M_\mathrm{SUSY}$ by using
the publicly available calculator \texttt{SOFTSUSY}
\cite{allanach2002}. The output is put into \texttt{micrOMEGAs}
\cite{belanger2011}, a publicly available code, which calculates a
number of phenomenological -- see \Sref{pheno} -- and cosmological
-- see \Sref{cosmo} -- observables assisting us to restrict the
parametric space of our model.

\subsection{Phenomenological Constraints}
\label{pheno}

The phenomenological considerations which we take into account in
our study are the following:

\subsubsection{SM fermion masses}\label{pfm}

In order to determine
with good precision the running of the Yukawa coupling constants,
we have to take into account the sizable SUSY corrections
\cite{carena1994,hempfling1994,hall1994,pierce1997,carena2000} to
the tree-level $b$-quark and $\tau$-lepton masses. We incorporate
them in our code using the formulas of \cref{pierce1997} in
accordance with the recent reanalysis of
\cref{anandakrishnan2014}. The result is to be compared with the
experimental value of $m_b(M_Z)$. This is derived by using as an
input parameter the \MS value, which, at $1-\sigma$, is
\cite{particledata}
\begin{equation}
 m_b(m_b)^\text{\MS}=4.18\pm0.03\gev.
\end{equation}
This range is evolved up to $M_Z$ using the central value
$a_s(M_Z)=0.1185$ \cite{particledata} of the strong fine-structure
constant at $M_Z$ and then converted to the \DR scheme in
accordance with the analysis of \cref{tobe2003}. We obtain
\begin{equation}
 \label{mbr}  2.8\lesssim m_b(M_Z)/\GeV\lesssim2.86
\end{equation}
at $68\%$ c.l. with the central value being $m_b(M_Z)=2.83\gev$.
Less important but not negligible (almost 4$\%$) are the SUSY
corrections \cite{pierce1997} to the $\tau$-lepton mass, which
have the effect \cite{gomez2002,gomez2003} of slightly reducing
$\tan\beta$. We take as input the central value \cite{tobe2003}
\begin{equation}
 m_\tau(M_Z)=1.748\gev
\end{equation}
of the \DR $\tau$-lepton mass at $M_Z$. Finally, as regards the
$t$-quark mass, we take the latest $1-\sigma$ result
\cite{mtnew} on its pole mass ($M_t=173.34\pm0.76\gev$) and
construct the 68\% c.l. range for its running mass $m_t(m_t)$:
\begin{equation}
 \label{mtr} 164.1\lesssim m_t(m_t)/\GeV\lesssim165.56
\end{equation}
with the central value being $m_t(m_t)=164.83\gev$.

\subsubsection{The mass of the lightest Higgs boson}\label{pmh}

The experiments ATLAS and CMS in the LHC discovered
simultaneously a boson that looks very much like the expected SM
Higgs boson. Its measured mass turns out to be
\cite{Aad:2014aba,CMS:2014ega}
\begin{equation}\label{eq:higgsmass}
  m_h=\begin{cases}
  125.36\pm0.37~(\text{stat.})\pm0.18~(\text{sys.})\gev~(\text{ATLAS}),\\
  125.03_{-0.27}^{+0.26}~(\text{stat.})_{-0.15}^{+0.13}~(\text{sys.})
    \gev~(\text{CMS}).
     \end{cases}
\end{equation}
While there is no combined analysis of the two experiments yet, we
can estimate the allowed $95\%$ c.l. range of this mass including
a theoretical uncertainty of about $\pm1.5\gev$. This gives
\begin{equation}
\label{1mhb}  122\lesssim m_h/\GeV\lesssim128.5.
\end{equation}
In most of the plots in \Sref{res}, we set
\begin{equation}\label{mhb}
  \mh=125.5\gev,
\end{equation}
which lies well inside the experimental range from both ATLAS and
CMS in Eq.~(\ref{eq:higgsmass}).

\subsubsection{Sparticles searches} \label{plhc}

The mass limits which are relevant for our investigation are the
following:

\paragraph{Mass of the chargino}\label{pch}
The combined results of experiments that took place at LEP
\cite{LEPchargino} showed that, regardless of the model, the lower
bound on the mass of the charginos is
\begin{equation}\label{mchb}
  \mch\gtrsim103.5\gev.
\end{equation}

\paragraph{Mass of the gluino}\label{pgl}
The present model dependent $95\%$ c.l. lower bound on the mass
of the gluino is \cite{ATLAScol2013}
\begin{equation}\label{mglb}
  \mgl\gtrsim1.3\tev.
\end{equation}
It is well-known that such a heavy gluino spoils the
success of several models which consider the HB/FP region
\cite{abe2007,horton2010,younkin2012,abe2012,gogoladze2013,
yanagida2013,yanagida2013b,delgado2014}. However, as we will
show, this does not happen in our case.

\subsubsection{$B$-physics constraints} \label{pbph}

We also consider the following constraints originating from
$B$-meson physics:

\begin{itemize}

\item The branching ratio $\bsmm$ of the process
$B_s\to\mu^+\mu^-$ \cite{bsmm, mahmoudi} is to be consistent with
the 95\% c.l. experimental upper bound
\cite{LHCb2012,albrecht2012}
\begin{equation}\label{bmmb}
 \mathrm{BR}(\bsmumu)\lesssim4.2\times10^{-9}.
\end{equation}
Although the most recent experimental result is
\cite{LHCb2013}
\begin{equation}
  1.1\lesssim\mathrm{BR}(\bsmumu)/10^{-9}\lesssim6.4,
\end{equation}
we use the more stringent upper bound in \Eref{bmmb}, since it is
a combined result and, thus, much more reliable.

\item The branching ratio $\bsg$ of $b\to s\gamma$
\cite{microbsg,nlobsg,nlobsg1,nlobsg2} must lie in the $95\%$ c.l.
range \cite{bsgexp, bsgSM, bsgSMa}
\begin{equation}
  2.79\times 10^{-4}\lesssim \bsg \lesssim 4.07\times 10^{-4}.
\label{bsgb} \end{equation}

\item The ratio $\btn$ of the CMSSM to the SM branching ratio of
the process $B_u\to \tau\nu$ \cite{mahmoudi,Btn} should be
confined in the $95\%$ c.l. range \cite{bsgexp}
\beq 0.52\lesssim\btn\lesssim2.04\,.\label{btnb} \eeq

\end{itemize}

\subsubsection{Muon anomalous magnetic moment}\label{pam}
There is a discrepancy $\damu$ between the measured value of the
muon anomalous magnetic moment $a_\mu$ from its SM prediction. The
latter, though, is not yet completely stabilized mainly because of
the ambiguities in the calculation of the hadronic
vacuum-polarization contribution. According to the evaluation of
this contribution in \cref{davier2011}, there is still a
discrepancy between the findings based on the $e^+e^-$-annihilation
data and the ones based on the $\tau$-decay data. However, in
\cref{jegerlehner2011}, it is claimed that this discrepancy can be
alleviated. Taking into account the recent and more reliable
calculation based on the $e^+e^-$ data \cite{hagiwara2011}, the
complete tenth-order QED contribution \cite{aoyama2012}, and the
experimental measurements \cite{muong2col2006} of $a_\mu$, we end
up with a $2.9-\sigma$ discrepancy
\beqs\begin{equation}\label{1damub}
  \damu=(24.9\pm8.7)\times10^{-10},
\end{equation}
resulting to the following 95\% c.l. range:
\begin{equation}\label{damub}
  7.5\times10^{-10}\lesssim\damu\lesssim42.3\times10^{-10}.
\end{equation}\eeqs
This $\delta a_\mu$ can be attributed to SUSY contributions
calculated by using the formulas of \cref{martin2001}. The resulting
$\damu$ has the sign of $\mu$ and its absolute value decreases as
$\mx$ increases. Therefore, \Eref{damub} hints that the sign of
$\mu$ has to be positive. Moreover, a lower [upper] bound on $\mx$
can be derived from the upper [lower] bound in \Eref{damub}. As it
turns out, only the upper bound on $m_\mathrm{LSP}$ is relevant
here. Taking into account the aforementioned computational
instabilities and the fact that a discrepancy at the level of
about $3-\sigma$ cannot firmly establish a real deviation from the
SM value, we restrict ourselves to just mentioning at which level
\Eref{1damub} is satisfied in the parameter space allowed by all
the other constraints -- cf.
Refs.~\cite{Ellis2012,baer2012,Buchmueller:2013rsa,ros15}.

\subsection{Cosmological  Constraints} \label{cosmo}

Our cosmological considerations include information from the CDM
abundance and direct detection experiments quoted below.

\subsubsection{Cold dark matter abundance} \label{pcdm}

In accordance with the recently reported results
\cite{Ade:2013zuv} from the Planck satellite, the 95\% c.l. range
for the CDM abundance in the universe is
\begin{equation}\label{eq:omegalimit}
 \cdm=0.1199\pm0.0054.
\end{equation}
In the context of the CMSSM, the lightest neutralino $\xx$ can be
the LSP and, thus, naturally arises as a CDM candidate as long as
its relic abundance does not exceed the upper bound in
Eq.~(\ref{eq:omegalimit}), i.e.
\begin{equation}\label{omgb}
  \omg\lesssim0.125.
\end{equation}
This is a quite strong restriction on the parameter space, since
$\omg$ increases, in general, with $\mx$ and so an upper bound on
$\mx$ can be derived from Eq.~(\ref{omgb}). Note that no lower
bound on $\omg$ is imposed in our analysis, since the CDM may
receive contributions from other particles too
\cite{covi2004,choi2012,baer2008,baer2009,baer2010}.

\begin{figure}[t]
\centerline{\epsfig{file=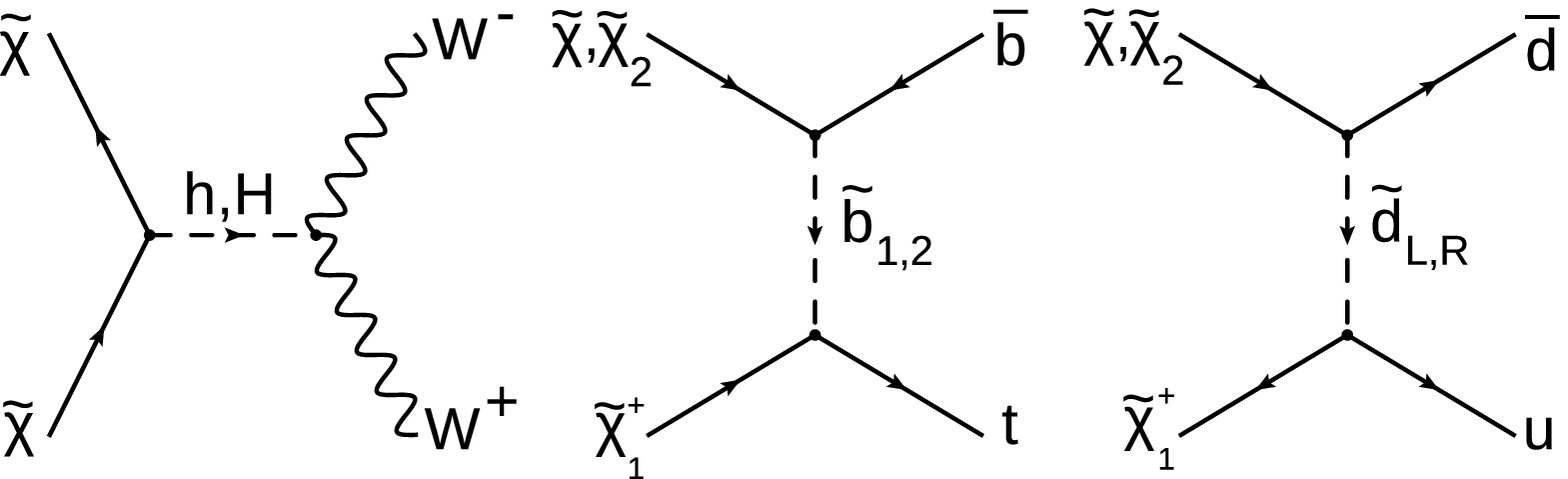,width=8.5cm}}
\caption{\footnotesize Dominant $\xx-\xx$ annihilation and
$\xx/\neutralino_2-\cha$ coannihilation reactions in the
HB/FP region.}\label{fig:hbfpannihilations}
\end{figure}

Focusing on the HB/FP region, we find that the main processes
causing the reduction of $\omg$ are the $\xx-\xx$ annihilations
(for large $|\amg|$'s and low $\mx$'s) and the $\xx/\xxb-\cha$
coannihilations (for low $|\amg|$'s and large $\mx$'s).
Specifically, the dominant reactions are (see
\Fref{fig:hbfpannihilations}) -- the notation used for the
various (s)particles is the same as in Table~\ref{tab2} --
\begin{itemize}
\item $\xx\xx\rightarrow W^+W^-$ mediated by a $h$ and $H$
exchange in the $s$-channel and
\item $\xx/\xxb~\cha\to t\bar b$ and $u\bar{d}$
mediated by a $\tilde{b}_{1,2}$ or $\tilde{d}_{L,R}$ exchange in
the $t$-channel, respectively.
\end{itemize}
Here, as in \Sref{resall}, we consider the squarks
$\tilde{d}_{L,R}$ of the two lightest generations as degenerate.
The strength of the coannihilation processes is controlled by
the relative mass splittings $\dch$ or $\dnt$ between the $\cha$
or $\xxb$, respectively, and the LSP, which are defined by
\beq \Delta_{\rm CA}=(m_{\rm CA}-\mx)/\mx,~~{\rm
CA}=\ch,\xxb\label{dchdef}.\eeq
In particular, we find that the resulting $\omg$ decreases with
$\dch$ or $\dnt$. The $\xx/\xxb-\cha$ coannihilation processes
are dominant in the regions where the bound on $\omg$ in
\Eref{omgb} is approached contributing almost $40\%$ to the
total effective cross section. On the other hand, in the region
where $\mch$ is near its lower limit in \Eref{mchb}, the
$\xx-\xx$ annihilation processes contribute more than $50\%$
to the reduction of $\omg$ -- cf. \cref{microfp}.

\subsubsection{Spin-independent direct detection of CDM}
\label{pssi}

Direct detection of dark matter through its elastic scattering off
atomic nuclei inside an underground detector would be an
undeniable evidence of its existence. Data coming from direct
detection experiments can provide strict bounds on the values of
the free parameters in SUSY models with a sizable higgsino
component, as in our case. Here, we focus on the \textit{large
underground Xenon} (LUX) experiment \cite{LUX2014}, whose
present data are a little more restrictive than those from the
XENON100 experiment \cite{xenon100}. Both belongs to the
experiments probing for the \textit{spin-independent} (SI) cross
section, since they consider only the scalar part of the cross
section. Therefore, the quantity which has to be considered for
comparing the experimental results and theoretical predictions is
the SI neutralino-proton ($\tilde\chi-p$) elastic scattering cross
section $\ssi$. This quantity is calculated by employing the
relevant routine of the {\tt micrOMEGAs} package \cite{Detmicro}
based on the full one-loop treatment of \cref{drees}. The scalar
form factors $f^p_{{{\rm T}q}}$ (with $q=u,d,s$) for light quarks
in the proton, which are needed for the calculation of $\ssi$, are
estimated following the revised approach of Ref.~\cite{Detmicro12}
and taking into account the recent lattice simulations
\cite{latticef}, which suggest the following $68\%$ c.l. ranges
for the pion-nucleon ($\sigma_{\pi N}$) and the
strangeness-nucleon ($\sigma_{s}$) sigma terms:
\beq \sigma_{\pi N} = 45\pm6~{\rm MeV} ~~\mbox{and}~~ \sigma_{s} =
21\pm6~{\rm MeV}.\eeq
Using also as inputs the light quark mass ratios
\cite{particledata} $m_u/m_d=0.48\pm0.1$ and $m_s/m_d=19.5\pm2.5$,
we find the following $1-\sigma$ ranges for the $f_{{\rm T}q}^p$'s:
\beqs\bea && \label{f1} f_{{\rm T}u}^p=0.018^{+0.0057}_{-0.0044},\\
   && \label{f2} f_{{\rm T}d}^p=0.026^{+0.0016}_{-0.0019},\\
  && f_{{\rm T}s}^p=0.022\pm0.0064. \label{f3}\eea\eeqs
Note that these ranges are much narrower than the ones used
in \cref{karagiannakis2011}. Also $f_{{\rm T}s}^p$ turns out to be
considerably smaller than its older value -- cf. \cref{profumo} --
reducing thereby the extracted $\ssi$.

In the HB/FP region, where the LSP has a significant higgsino
component, $\ssi$ is dominated by the $t$-channel Higgs-boson-exchange
diagram contributing to the neutralino-quark elastic scattering
process -- for the relevant tree-level interaction terms see e.g.
the appendix of \cref{Lazarides:2004pq}. Especially for large
$\tan\beta$'s, which is the case here, the couplings of the heavier
Higgs boson $H$ to down-type quarks are
proportional to $\tan\beta$ and so are the dominant ones. More
explicitly, $\ssi$ behaves as
\begin{equation}
 \label{ssiN} \ssi\propto \tan^2\beta|N_{1,1}|^2|N_{1,3}|^2/m_H^4,
\end{equation}
where $N_{1,1}$, $N_{1,2}$, and $N_{1,(3,4)}$ are the elements of
the matrix $N$ which diagonalizes the neutralino mass matrix and
express the bino, wino, and higgsino component of $\xx$,
respectively. As a consequence, $\ssi$ can be rather enhanced
compared to its value in the $\xstau$ coannihilation region, where
$\xx$ is essentially a pure bino.

The data from the LUX experiment, which we take from \cref{brown},
are directly applicable in the case where the CDM consists of just
one kind of weakly interacting massive particle, which in our case
would be the neutralino $\xx$. However, if the $\xx$'s constitute
only a part of the CDM in the universe, some extra care is
required. The number of the scattering events
$\lambda$ is \cite{Lewin1996,green2007,green2008} proportional to
$\ssi$ and the local $\xx$ density $\rho_{\tilde\chi}$:
\begin{equation}\label{eq:lambda}
  \lambda\propto\ssi\rho_{\tilde\chi}.
\end{equation}
In the case where the CDM consists only of neutralinos,
\begin{equation}\label{eq:lambdaCDM}
      %\lambda\propto\sigma_{CDM-p}^{\mathrm{SI}}\rho_{\mathrm{CDM}},
    \rho_{\tilde\chi}=\rho_{\mathrm{CDM}},
\end{equation}
where $\rho_{\mathrm{CDM}}$ is the local CDM density. So, the LUX
experiment bound on $\lambda$ can be directly translated into a
bound on $\ssi$ provided that $\rho_{\mathrm{CDM}}$ can be
estimated -- see e.g. Ref.~\cite{Catena:2009mf}. However, when the
$\xx$'s constitute only a fraction of the total CDM, we can write
Eq.~(\ref{eq:lambda}) as
\begin{equation}\label{eq:lambdaneutralino}
\lambda\propto\ssi\frac{\rho_{\tilde\chi}}{\rho_{\mathrm{CDM}}}
\rho_{\mathrm{CDM}}.
\end{equation}
Following the authors of Ref.~\cite{bertone2010}, we can then use
the scaling ansatz
\begin{equation}\label{eq:scalingansatz}
  \rho_{\neutralino}/\rho_{\mathrm{CDM}}=\Omega_{\neutralino}/
    \Omega_{\mathrm{CDM}},
\end{equation}
which gives
\begin{equation}
  \lambda\propto\sigma_{\tilde{\chi}p}^{\mathrm{SI}}
    \frac{\Omega_{\tilde{\chi}}}{\Omega_{\mathrm{CDM}}}
    \rho_{\mathrm{CDM}}.
\end{equation}
So, the LUX experiment bound on $\lambda$ is now translated into a
bound on the `rescaled' SI neutralino-proton elastic cross section
$\xi\sigma_{\tilde{\chi}p}^{\mathrm{SI}}$, where $\xi=
\Omega_{\tilde{\chi}}/\Omega_{\mathrm{CDM}}$.

\section{The hyperbolic branch/focus point region} \label{hbfp}

A detailed discussion of the characteristics of the HB/FP
region of the CMSSM is given in Refs.~\cite{akula2012,nath13}. The
classification of the various solutions of the radiative EWSB
condition is based on the expansion of $\mu^2$ in terms of the soft
SUSY breaking parameters of the CMSSM included in \Eref{par}.
Indeed, using fitting techniques, we can verify the following
formula
\beq\mu^2+M_Z^2/2\simeq c_0
m^2_0+c_{1/2}M^2_{1/2}+c_{A}A_0^2+c_{AM}A_0\mg,
\label{mufit}\eeq
where the coefficients $c_0, c_{1/2}, c_A$, and $c_{AM}$ depend
basically on $\tnb$ and the masses of the fermions of the third
generation. These coefficients are computed at the scale
$M_{\rm SUSY}$ in \Eref{opt} and, therefore, inherit a mild
dependence on the SUSY spectrum too. From \Eref{mufit}, we can
easily infer that the SUSY breaking parameters are bounded above
for fixed $\mu$, when the quadratic form in the right-hand side
of this equation is positive definite. This is the so-called
\textit{ellipsoidal branch} (EB). On the contrary, in the
\textit{hyperbolic branch} (HB) region \cite{chan1998}, $c_0$
is negative and, consequently,
$m_0$ can become very large together with a combination of $A_0$
and $\mg$ with all the other parameters being fixed. Near the
boundary between the EB and HB regions, the coefficient $c_0$ is
very close to
zero and, thus, $m_0$ can become very large with all the other
parameters fixed. Moreover, there is a region where
the soft SUSY breaking mass-squared $m_{H_2}^2$ of
$H_2$ becomes independent of the asymptotic value of the
parameter $m_0$. This is called the \textit{focus point} (FP)
region \cite{feng, feng12, focus2014}.

In the HB region, the radiative EWSB admits three types of
solutions
\cite{akula2012,nath13}:
\begin{enumerate}

\item[(i)]  Focal Points (HB/FP): They lie near the  boundary
between the EB and the HB regions, where $c_0\simeq0$ and, thus,
$\mu^2$ is practically independent of $m_0^2$. As a consequence,
$m_0^2$ can become very large, while $\mu^2$ and all the other
parameters remain fixed. We should note that, in the large $\tnb$
regime, we obtain $\mu^2+M_Z^2/2\simeq-m_{H_2}^2$ and so the focal
points, where $\mu^2$ is essentially independent of $m_0^2$,
coincide with the focus points, where $m_{H_2}^2$ is almost
$m_0^2$-independent. Since, in the present model, $\tnb$ is large,
we will not distinguish focal from focus points and we will use
the same abbreviation (FP) for both.

\item[(ii)] Focal Curves (HB/FC): Along these curves, two of the
soft parameters can acquire large values, while $\mu^2$ and all
the other parameters remain fixed.

\item[(iii)] Focal Surfaces (HB/FS): Here we can have three soft
parameters with large values, while $\mu^2$ and the other
parameters remain fixed.
\end{enumerate}

\begin{table}[t] \caption{The $c$'s in \Eref{mufit} for the
four cases of \Tref{tab2}.}
\begin{ruledtabular}
\begin{tabular}{c|ccccc}
$\tnb$&  $c_0$  &  $\delta c_0$  &$c_{1/2}$&$c_{A}$&$c_{AM}$\\
\hline\\ [-0.2cm]
$40$&  $-0.0921$&$0.014$ & $0.789$& $0.107$  &$-0.269$ \\
$45$&  $-0.0775$&$0.0148$ & $0.825$& $0.1016$ &$-0.260$  \\
$48$& $-0.0686$&$0.0151$ & $0.845$& $0.0981$ &$-0.253$ \\
$50$&  $-0.0624$&$0.0265$ & $0.859$& $0.0953$ &$-0.247$  \\
\end{tabular}
\end{ruledtabular} \label{tab1}
\end{table}

To get an idea of how our solutions in \Sref{res} are classified
into these categories, we display, in \Tref{tab1}, the values
of the coefficients $c$ in \Eref{mufit} for the four
representative cases of \Tref{tab} corresponding to
$\tnb=40,45,48$, and 50 -- see \Sref{resall}. We also
list the relevant $1-\sigma$ uncertainly $\delta c_0$ of $c_0$,
which is practically the deviation between the lowest $c_0$
obtained by varying $m_t(m_t)$ in the $1-\sigma$ range in
\Eref{mtr} and the value of $c_0$ corresponding to the central
value of $m_t(m_t)$ -- possible variation of $m_b(M_Z)$ in its
$1-\sigma$ range in \Eref{mbr} does not change $c_0$ significantly.
As it turns out, the minimal $c_0$ corresponds to the upper limit
of $m_t(m_t)$ in \Eref{mtr}. Taking then the ratio
$|c_0|/\delta c_0$, we see that the displacement of $c_0$ from
zero ranges from $2.4-\sigma$ (for $\tnb=50$) to $6.6-\sigma$
(for $\tnb=40$). In particular, for $\tnb\gtrsim45$, this
displacement does not exceed $5-\sigma$ and, thus, a real
deviation of the coefficient $c_0$ from zero cannot be established. As a
consequence, we can say that the corresponding parameters lie
within the HB/FP region. On the contrary, for $\tnb\lesssim45$,
the parameters belong to the HB/FC and HB/FS region.

%Given that agreement with \Eref{1mhb} entails relatively large
%$m_0$'s it is easy to understand the attractiveness of this region
%for viable SUSY solutions. Therefore, in our solutions we expect $\mo$ and therefore the
%whole spectrum of sfermions, sleptons and heavier higgses to be
%rather high and more or less unconstrained.

\section{Restrictions on the SUSY Parameters}\label{res}

Imposing the requirements described in \Sref{sec:pheno} -- with
the exception of the one of \Sref{pam} --, we can restrict the
parameters of our model. Following our approach in
Refs.~\cite{karagiannakis2012,karagiannakis2013}, we use as free
parameters of the model the ones in \eq{par}. The Yukawa coupling
constant ratios $h_t/h_\tau$ and $h_b/h_\tau$ are then fixed by
using the data of \Sref{pfm}. These ratios must satisfy the
YQU conditions in \Eref{yquc} for natural values of the parameters
$\alpha_1,\alpha_2$, and $\rho$ -- see \Sref{yuk}. We restrict
ourselves to the range $40\leq\tnb\leq50$ since, below
$\tan\beta=40$, the ratios $h_m/h_n$ ($m,n=t,b,\tau$) tend to
require unnatural values of $\alpha_1,\alpha_2$, and $\rho$ and,
above $\tan\beta=50$, the numerical calculations for the soft SUSY
masses become quite unstable. As a characteristic value in our
allowed parameter space, we take $\tan\beta=48$, which balances
well enough between maintaining natural values for the $h_m/h_n$'s
and satisfying the various requirements of \Sref{sec:pheno}.

We concentrate on the $\mu>0$ case, given that $\mu<0$ worsens
the violation of \Eref{1damub}, and scan the region
$-30\leq \amg\leq30$. We will hereafter use the central values
of the SM parameters $M_t$, $m_b(M_Z)$, $m_\tau(M_Z)$, and
$\alpha_s(M_Z)$ given in \Sref{pfm}. We find that the only
constraints which play a role are the lower bound on $m_h$ in
Eq.~(\ref{mhb}), the bounds on $\mch$ and $\mgl$ in
\eqs{mchb}{mglb}, the CDM bound in Eq.~(\ref{omgb}), and the
bound from the LUX experiment --  see \Sref{pssi}. In the
parameter space allowed by these requirements,
all the other restrictions of Sec.~\ref{pheno} are automatically
satisfied with the exception of the lower bound on $\delta a_\mu$
in Eq.~(\ref{damub}). This bound is not imposed here as a strict
constraint on the parameters of the model for the reasons
explained in \Sref{pam}. We only discuss the level at which the
requirement in \Eref{1damub} is satisfied in the parametric
region allowed by all the other constraints.

We first present, in \Sref{resm12m0}, the restrictions in the
$\mg-\mo$
plane for low values of $\amg$ and then, in \Sref{resmgl}, we
reveal the role of the bound in \Eref{mglb} by considering the
$\tan\beta=48$ case and ignoring the restrictions of \Sref{pssi}.
In Secs.~\ref{reslux} and \ref{resall}, we present regions for
various $\tnb$'s consistent with all the requirements of
\Sref{sec:pheno} but the one of \Sref{pam}.

\subsection{\boldmath Restrictions in the $\mg-\mo$ Plane}\label{resm12m0}

\begin{figure}[t]
\centerline{\epsfig{file=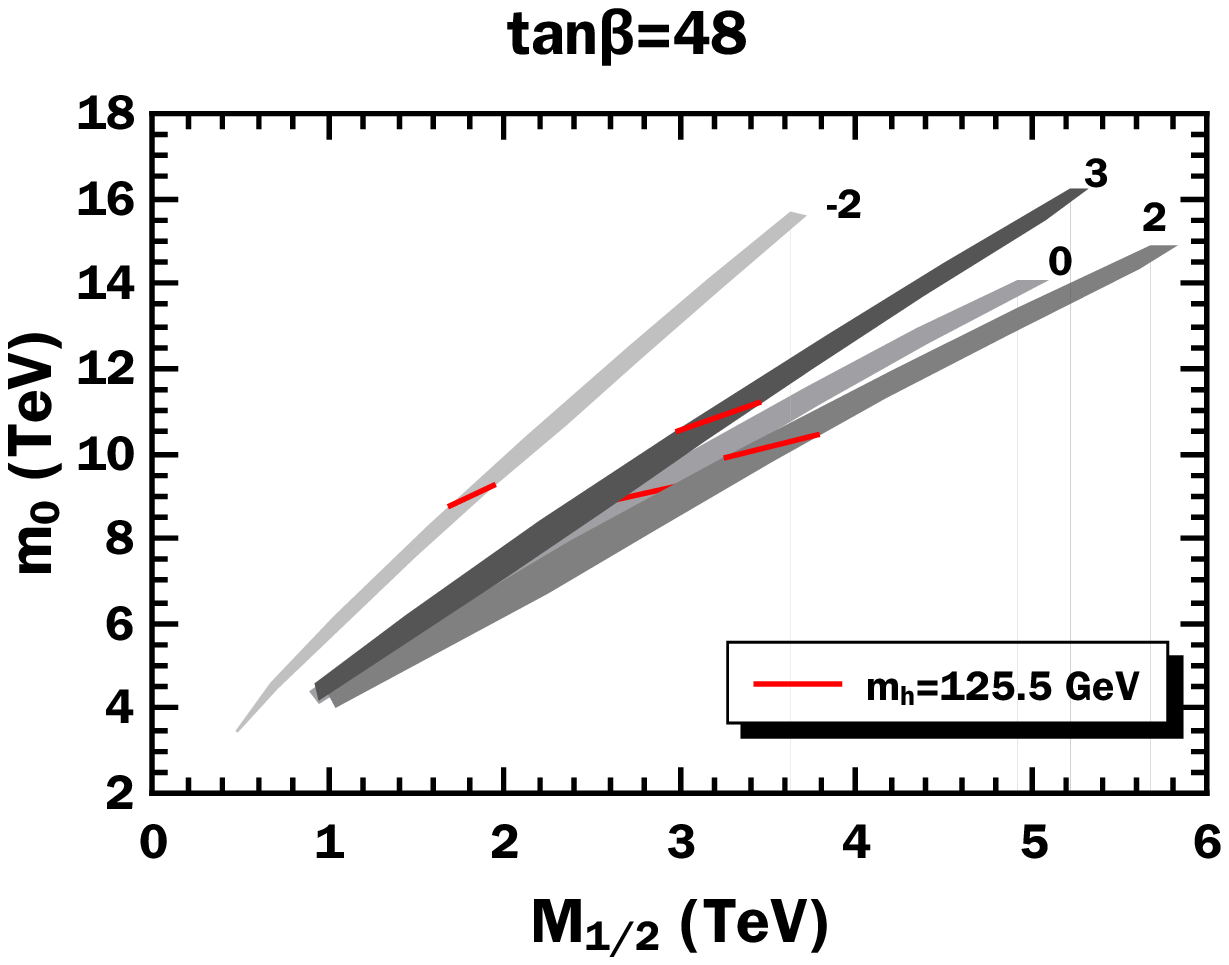,width=8.5cm}}
\caption{Allowed (shaded) regions in the $M_{1/2}-m_0$ plane for
$\tan\beta=48$ and different $\amg$'s indicated in the plot. The
red lines correspond to $\mh$ in Eq.~(\ref{mhb}).
}\label{fig:tanb48m12m0comp}
\end{figure}

The interplay of the various requirements of \Sref{sec:pheno} can
be easily understood from \Fref{fig:tanb48m12m0comp}, where we
present the (shaded) strips in the $M_{1/2}-m_0$ plane allowed by
the restrictions of Secs.~\ref{pfm}, \ref{pmh}, \ref{plhc},
\ref{pbph}, and \ref{pcdm} for $\tan\beta=48$ and several values
of $\amg$ indicated in the graph. The upper boundary along each
of these allowed strips arises from the limit on $\mch$ in
Eq.~(\ref{mchb}). Note that this limit is more restrictive than
the limit for triggering radiative EWSB -- cf. \cref{mayes2013}.
The lower boundary along each strip is given by the upper limit
on $\omg$ in Eq.~(\ref{omgb}). On the other hand, the lower limit
on $\mh$ in \Eref{1mhb} causes the termination of the strips at
low values of $m_0$ and $M_{1/2}$, whereas their termination
at high values of $m_0$ is put in by hand in order to avoid
shifting the SUSY masses to very large values. The red lines
indicate solutions with $\mh=125.5\gev$ -- see \Eref{mhb}.

From this figure, we easily see the main features of the HB/FP
region: $\mo$ spans a huge range $(4-15)\tev$, whereas $\mg$
[$\mu$] remains relatively low $(1-6)\tev$ [$(0.1-1)\tev$] thanks
to the structure of the radiative EWSB in this region analyzed in
\Sref{hbfp}. We observe also that as $\amg$ increases from $-2$
to $2$ the allowed strip moves to larger $M_{1/2}$'s and becomes
less steep, while, for $\amg>2$, we have exactly the opposite
behavior: as $\amg$ increases, the allowed strip moves to smaller
values of $M_{1/2}$ and, at the same time, becomes steeper.

\begin{figure}[t]
\centerline{\epsfig{file=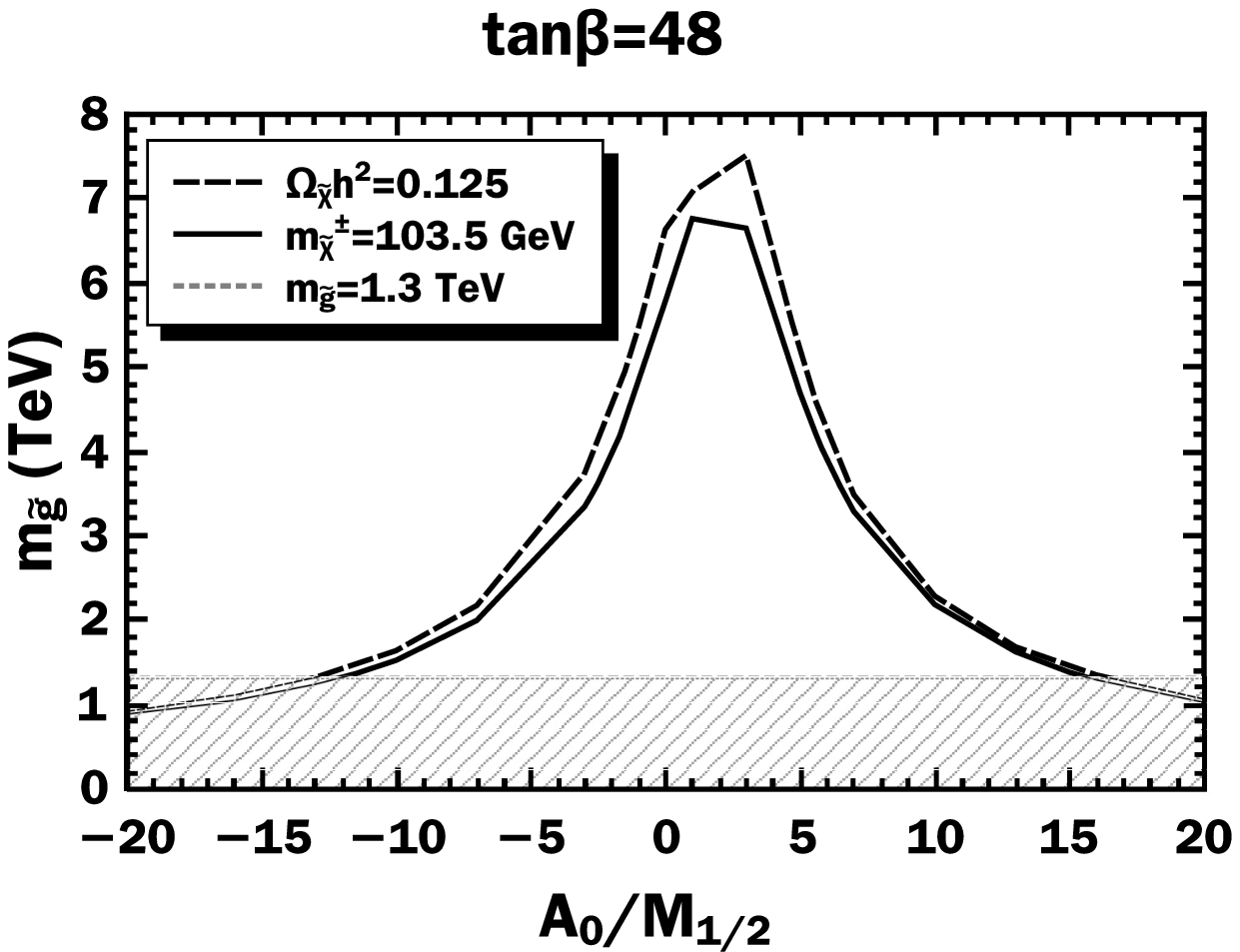,width=8.5cm}}
\caption{Restrictions in the $\amg-\mgl$ plane from
\eq{mchb} (solid line) and \eq{omgb} (dashed line) for $m_h$
in \Eref{mhb} and $\tnb=48$. The region excluded by
\Eref{mglb} is cross hatched.}\label{fig:a0gluino}
\end{figure}

\subsection{\boldmath Restrictions from the Bound on $\mgl$}\label{resmgl}

As $|\amg|$ increases above $10$, the restriction on $\mgl$ in
\Eref{mglb} comes into play. Its importance can be easily inferred
from \Fref{fig:a0gluino}, where we plot $\mgl$ against $\amg$ for
$\mh$ as in \Eref{mhb} and $\tan\beta=48$ -- let us note that
varying $\tan\beta$ does not significantly affect this figure. The
solid [dashed] line corresponds to the limit on $\mch$ [$\omg$] in
\Eref{mchb} [\Eref{omgb}]. The dotted line denotes the lower limit
on $\mgl$ in Eq.~(\ref{mglb}) excluding the cross hatched region.
Therefore, the allowed region (still without considering the LUX
data -- see \Sref{pssi}) is the region bounded by the three
aforementioned lines. We see clearly that the bound on $\mgl$ cuts
off the largest values of $|\amg|$. Specifically, near the lower
bound on $\mch$ given in Eq.~(\ref{mchb}), $\amg$ is limited in
the range
$-10.8\lesssim\amg\lesssim14.4$, while, near the upper bound on
$\omg$ in Eq.~(\ref{omgb}), it is limited  in the range
$-11.7\lesssim\amg\lesssim15.5$. In order to understand this
behavior, let us recall that $\mh$ depends on the ratio
${X_t^2}/m_{\sutop}^2$, where $X_t=A_t-\mu\cot\beta$ with $A_t$
being the soft trilinear scalar coupling constant for the
$t$-squark. For $\mh$ and $\tan\beta$ fixed, we see that larger
absolute values of $\amg$ require larger $m_{\sutop}$'s too and,
since $m_{\sutop}$ depends largely on $\mo$, the latter must grow
larger as well together with the left-handed squark soft SUSY
breaking masses $M_{Q_L}$. These soft masses increase as the
soft gluino mass
$M_3$ decreases -- see the relevant renormalization group equation
in \cref{barger1994} -- and, thus, it is quite obvious that, in
order to achieve larger values of $M_{Q_L}$, we need smaller
$M_3$'s, which also suggests smaller values for the common
asymptotic gaugino mass $M_{1/2}$. Therefore, small $M_{1/2}$'s
and large $|A_0|$'s, which lead to big $|\amg|$'s, yield
small
$\mgl$'s, which is exactly what one deduces from
\Fref{fig:a0gluino}.

\subsection{Restrictions from the LUX Experiment}\label{reslux}

\begin{figure*}[ht!]
\centerline{\epsfig{file=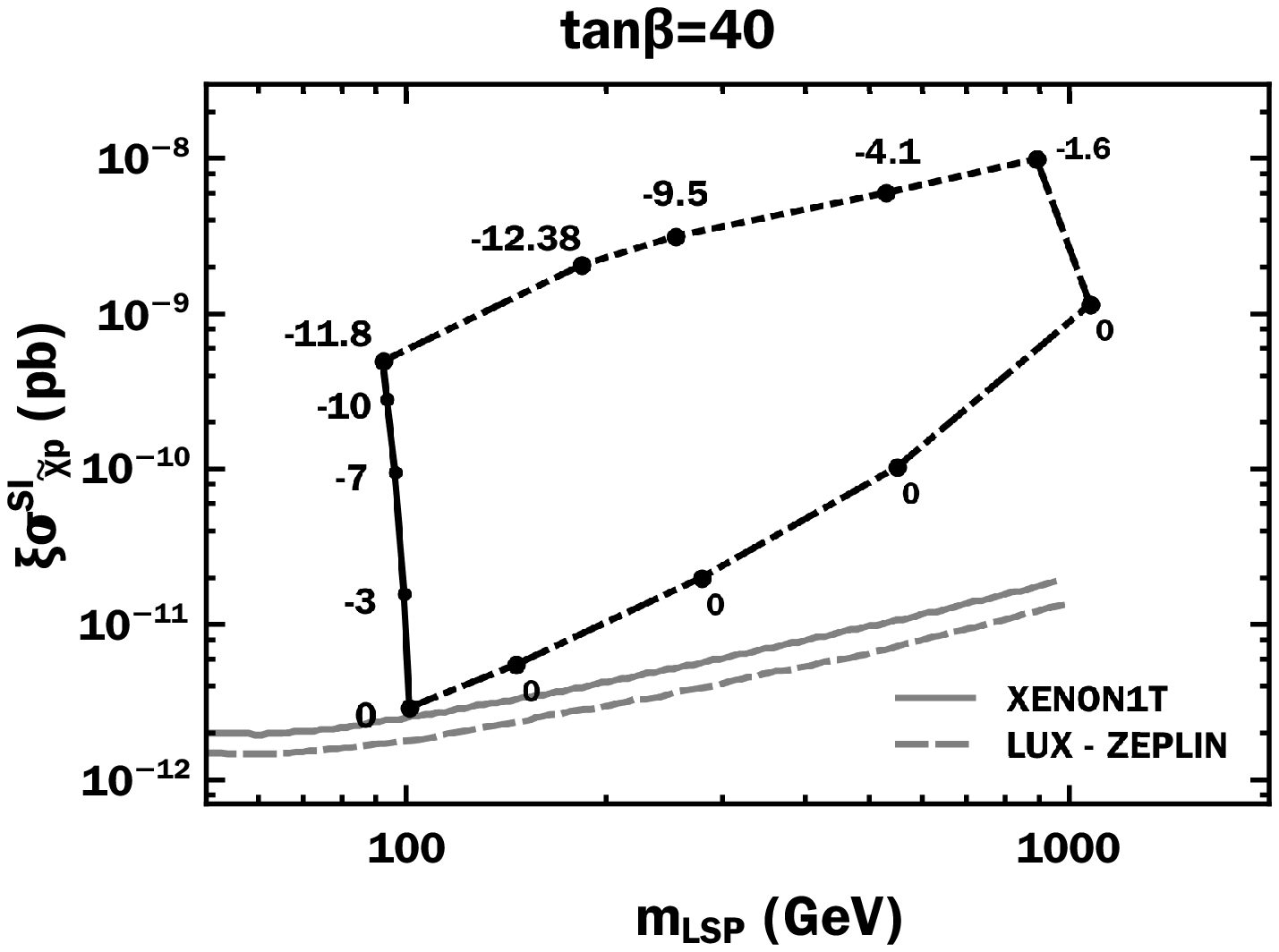,width=8.8cm}
\epsfig{file=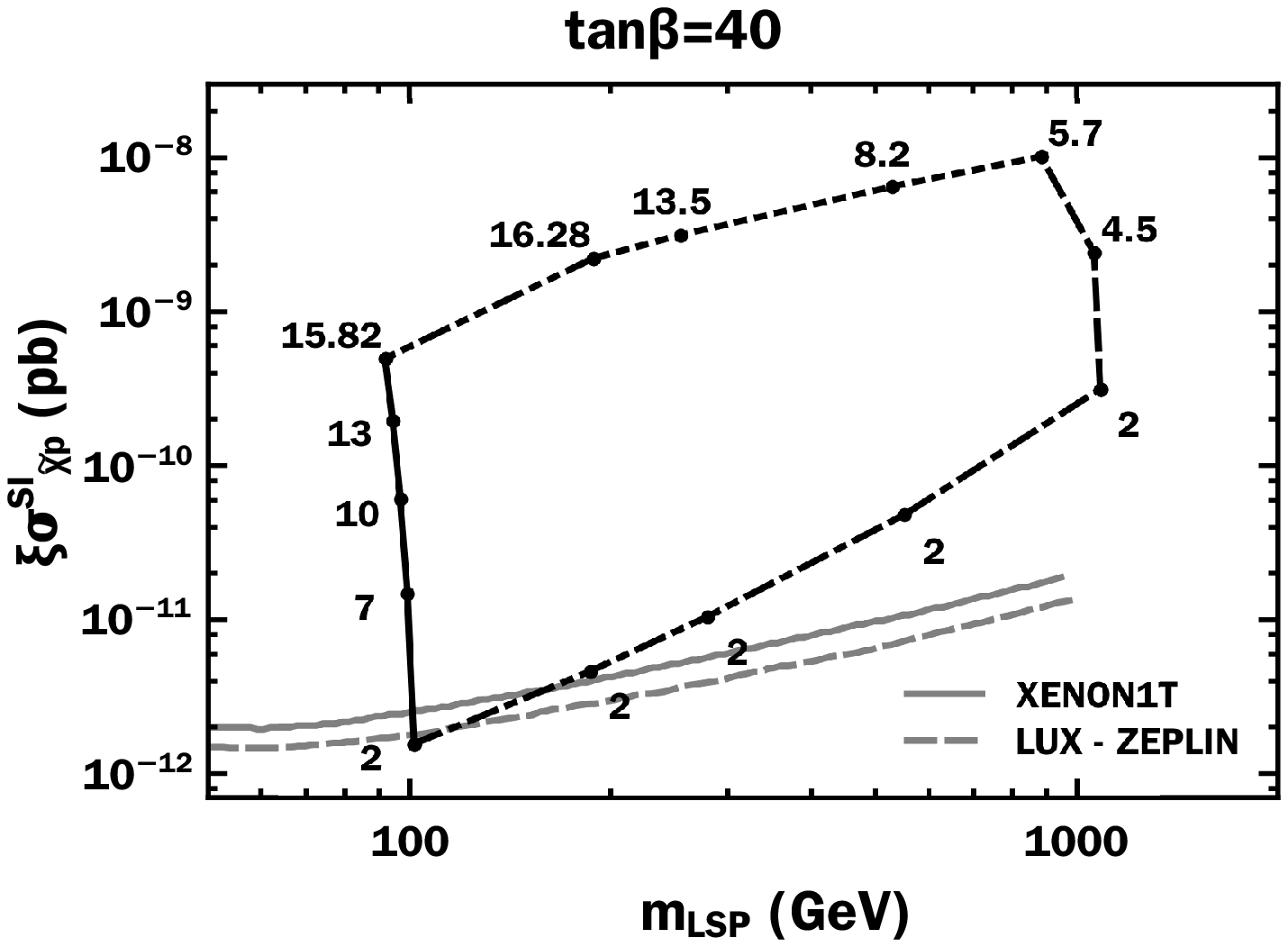,width=8.8cm}}\vspace*{1cm}
\centerline{\epsfig{file=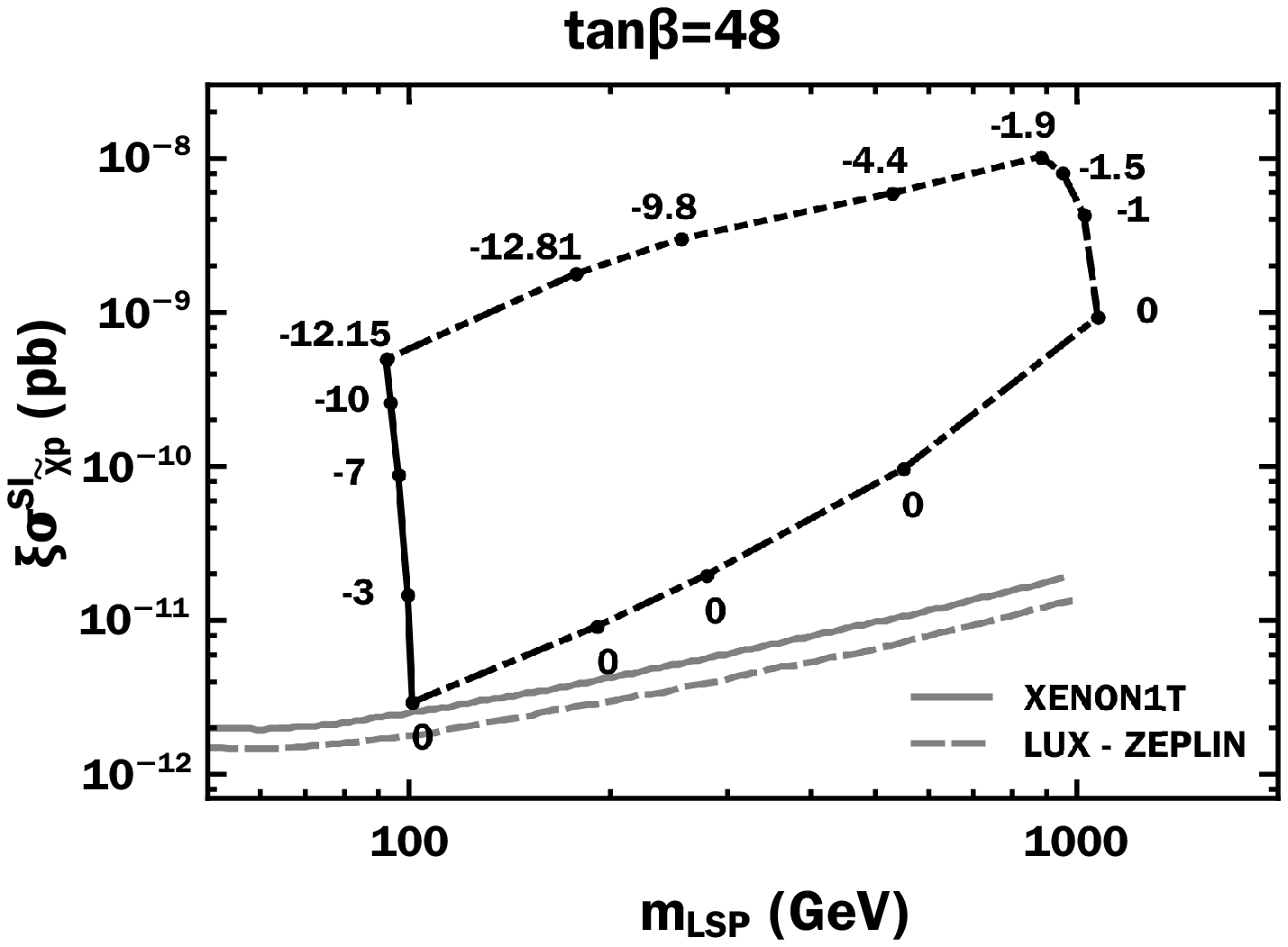,width=8.8cm}
\epsfig{file=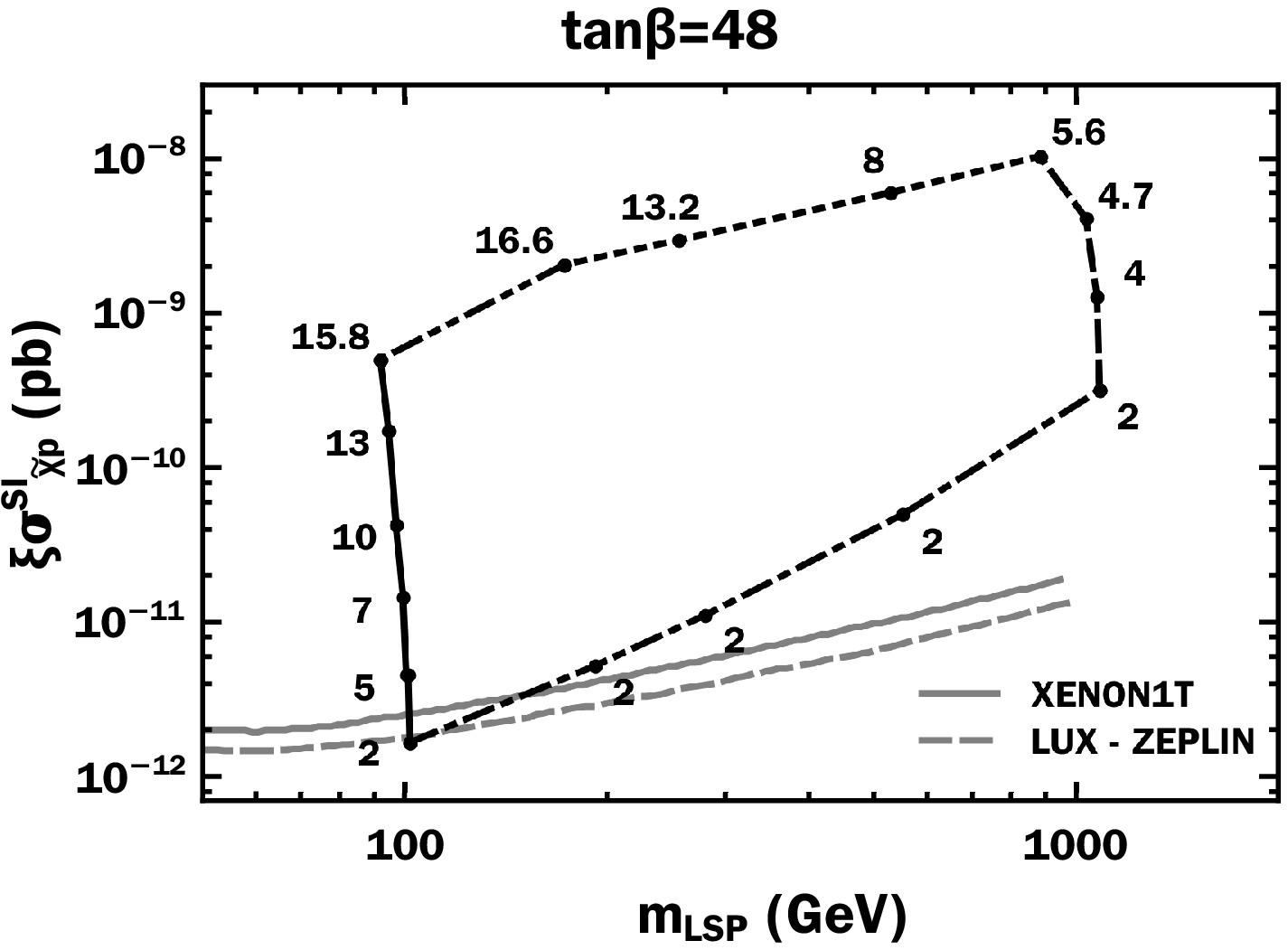,width=8.8cm}}\vspace*{1cm}
\centerline{\epsfig{file=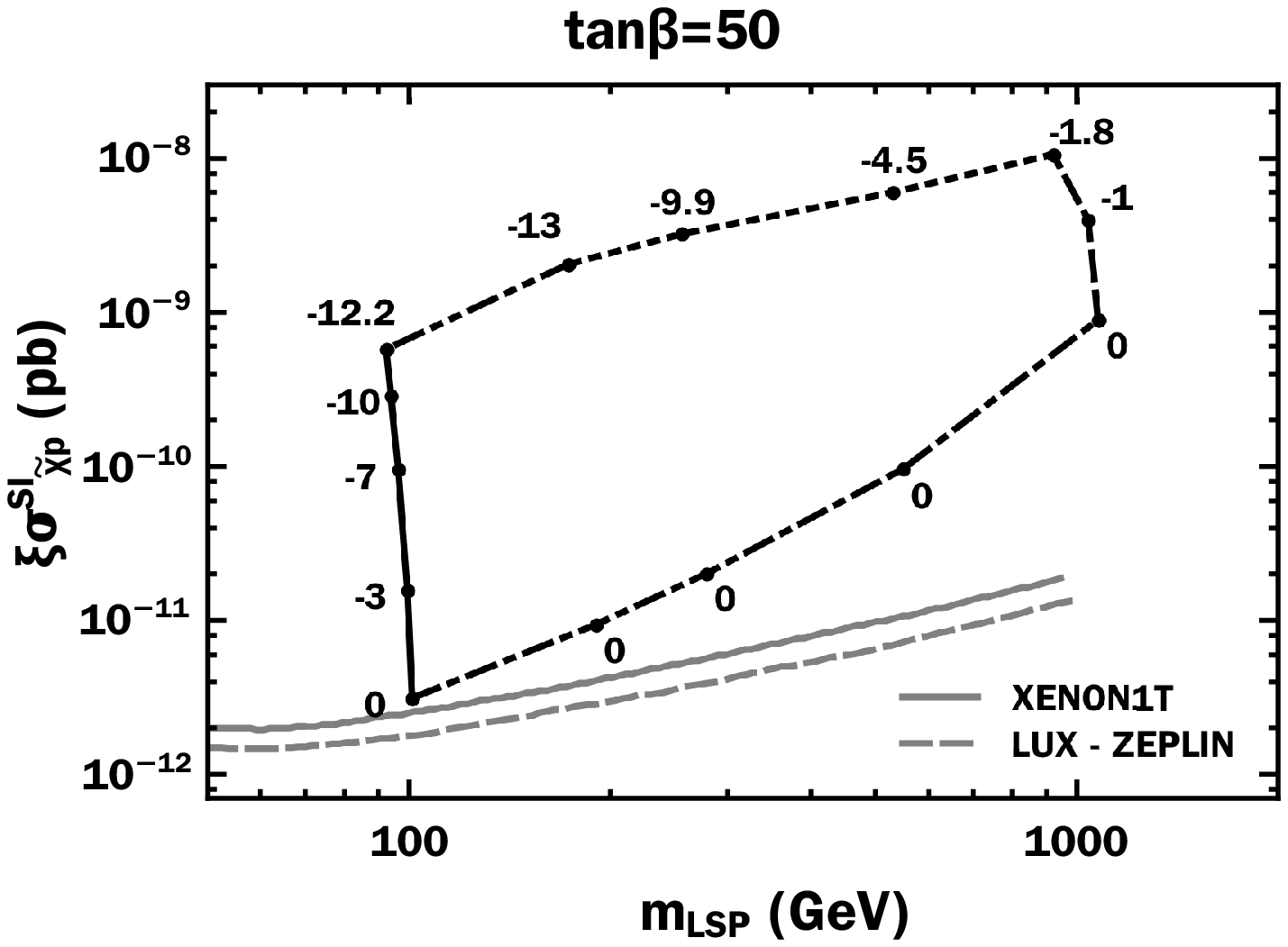,width=8.8cm}
\epsfig{file=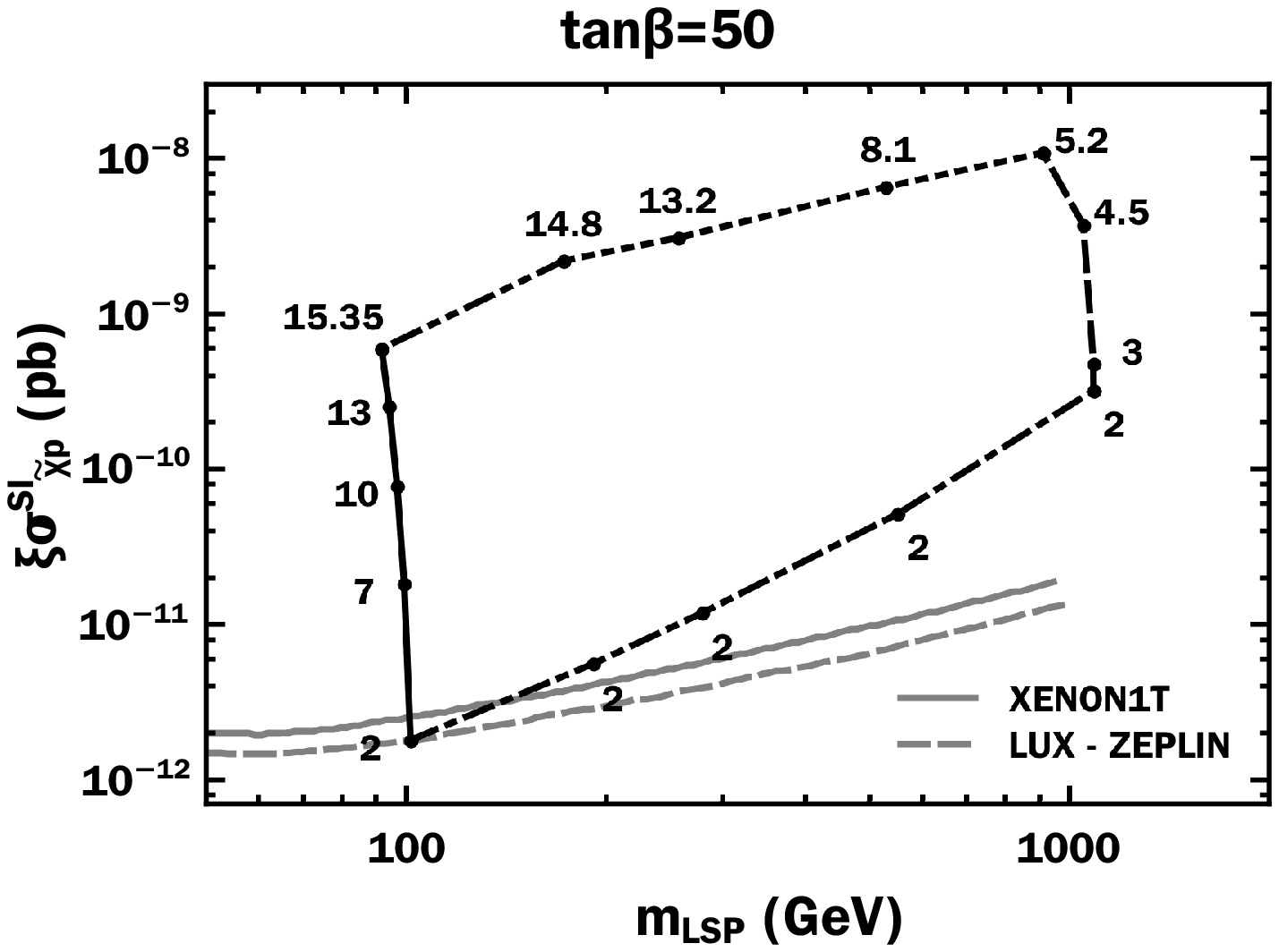,width=8.8cm}}\vspace*{1.cm}
\caption{Allowed (bounded) regions in the $\mx-\xssi$ plane for
$m_h$ given by \Eref{mhb} and $\tan\beta=40, 48$, and $50$. The
left [right] panels correspond to $\amg\leq 0$ [$\amg\geq 0$] and
the values of $\amg$ at the various points of the boundary lines
are indicated. The black solid, double dot-dashed, and dashed
lines correspond to the bounds from Eqs.~(\ref{mchb}),
(\ref{mglb}), and (\ref{omgb}), respectively. The black dotted
lines arise from the LUX data, whereas the black dot-dashed lines
give the lowest possible $\xssi$ in each case. The planned
sensitivity limits of XENON1T and LUX-ZEPLIN are also depicted by
a gray solid and dashed line, respectively.} \label{fig:xisigma}
\end{figure*}

Considering the cross section $\ssi$, we can not only probe the
detectability of the LSP, but also obtain further restrictions on
the parameters of our model. This is because of the rather
enhanced $\ssi$'s obtained in the HB/FP region, as explained in
\Sref{pssi}. In the computation of $\ssi$, we adopt the central
values of the $f_{{\rm T}q}^{p}$'s in Eqs.~(\ref{f1})-(\ref{f3})
and fix $\mh$ to its value in \Eref{mhb}. The results are
presented in Fig.~\ref{fig:xisigma}, where we depict the allowed
(bounded) regions in the $\mx-\xssi$ plane for $\tnb=40$ (upper
panels) $\tnb=48$ (middle panels) and $\tnb=50$ (lower panels).
We also draw with gray solid and dashed lines the projected
sensitivities \cite{brown} of the XENON1T \cite{xenon1t} and
LUX-ZEPLIN \cite{supercdms} experiments, respectively. The panels
in the left [right] column correspond to $\amg\leq0$
[$\amg\geq0$]. We see that, for each $\tnb$, the allowed regions
in the left and right panel almost coincide and, thus, we display
them separately to avoid confusion. The numbers on the various
points of each boundary line indicate the corresponding values of
$\amg$.

The allowed regions are bounded by five different types of
black lines for which we adopt the following conventions:
\begin{itemize}
\item on the solid line the bound on $\mch$ coming from
\Eref{mchb} is saturated;

\item on the double dot-dashed line the limit on $\mgl$ in
\Eref{mglb} is saturated;

\item on the dashed line the bound on $\omg$ from \Eref{omgb} is
saturated;

\item the dotted line depicts the bound on $\xssi$ arising from
the LUX data taken from \cref{brown};

\item the dash-dotted line represents the lowest possible
$\ssi$ in each case.
\end{itemize}
As can be understood from the description of the various lines
above, we have $\xi=1$ along the dashed line. As regards the
minimal $\xi$, this reaches $0.013$ independently of $\tnb$.

Notice that there are no significant differences between the
various $\tan\beta$'s as regards the allowed ranges of $\mx$ and
$\amg$. From all these graphs, we deduce that, as we move towards
the $\omg=0.125$ line, $\mx$ and $\mch$ grow larger, while the
allowed range of $\amg$ becomes smaller. The largest $|\amg|$ is
located at the junction point of the double dot-dashed and
dotted lines -- with the exception of the allowed region for
$\tnb=50$ and $\amg>0$, where the largest $|\amg|$ is found at the
intersection of the solid and double dot-dashed lines. The
smallest $\mx$ can be found in the upper left corner of the
allowed regions, whereas the largest $\mx$ is found on the dashed
line. The allowed ranges of $\amg$ and $\mx$ can be summarized as
follows:

\begin{itemize}
\item For $\tan\beta=40$, we find $-12.4\lesssim\amg\lesssim16.28$ and
$92\lesssim{\mx/\GeV}\lesssim1084.8$.

\item For $\tan\beta=48$, we find $-12.8\lesssim\amg\lesssim15.8$ and
$92\lesssim{\mx/\GeV}\lesssim1084.2$.

\item For $\tan\beta=50$, we find $-13\lesssim\amg\lesssim15.35$ and
$91.9\lesssim{\mx/\GeV}\lesssim1088$.

\end{itemize}

We noticed that, as $\amg$ approaches $2$ from above, the value
of the product $|N_{1,1}|^2|N_{1,3}|^2$ decreases. However, as
$\amg$ decreases below $2$, $|N_{1,1}|^2|N_{1,3}|^2$ grows. The
growth of this product persists even when $\amg$ becomes
negative.  As a consequence -- see \Eref{ssiN} -- $\xssi$
acquires its minimal value at $\amg=2$. The overall minimum of
$\xssi$, for each value of $\tnb$, is acquired at the lowest
left corner of the corresponding allowed region in the right
column of \Fref{fig:xisigma}. The smallest of these overall
minima, for $\tnb\geq 40$, is
\beq \xssi\gtrsim 1.56\times10^{-12}~{\rm
pb}~\lf1.49\times10^{-12}~{\rm pb}\rg, \label{sgmxp}\eeq
and it is obtained at $\tnb=40$ and $\mx\simeq101.7\gev$. The
bound in parenthesis is derived by allowing the $f_{{\rm
T}q}^{p}$'s to vary within their $1-\sigma$ intervals in
Eqs.~(\ref{f1}), (\ref{f2}), and (\ref{f3}). As shown in the
plots, the obtained values of $\xssi$ are within the reach of
forthcoming experiments like XENON1T and LUX-ZEPLIN with planned
sensitivity from $10^{-45}$ to $10^{-47}~{\rm cm}^2$ for the $\mx$
range under consideration -- recall that
$1~\mathrm{pb}=10^{-36}~{\rm cm}^{2}$.

\subsection{The Overall Allowed Parameter Space} \label{resall}

\begin{figure}[t!]
\centerline{\epsfig{file=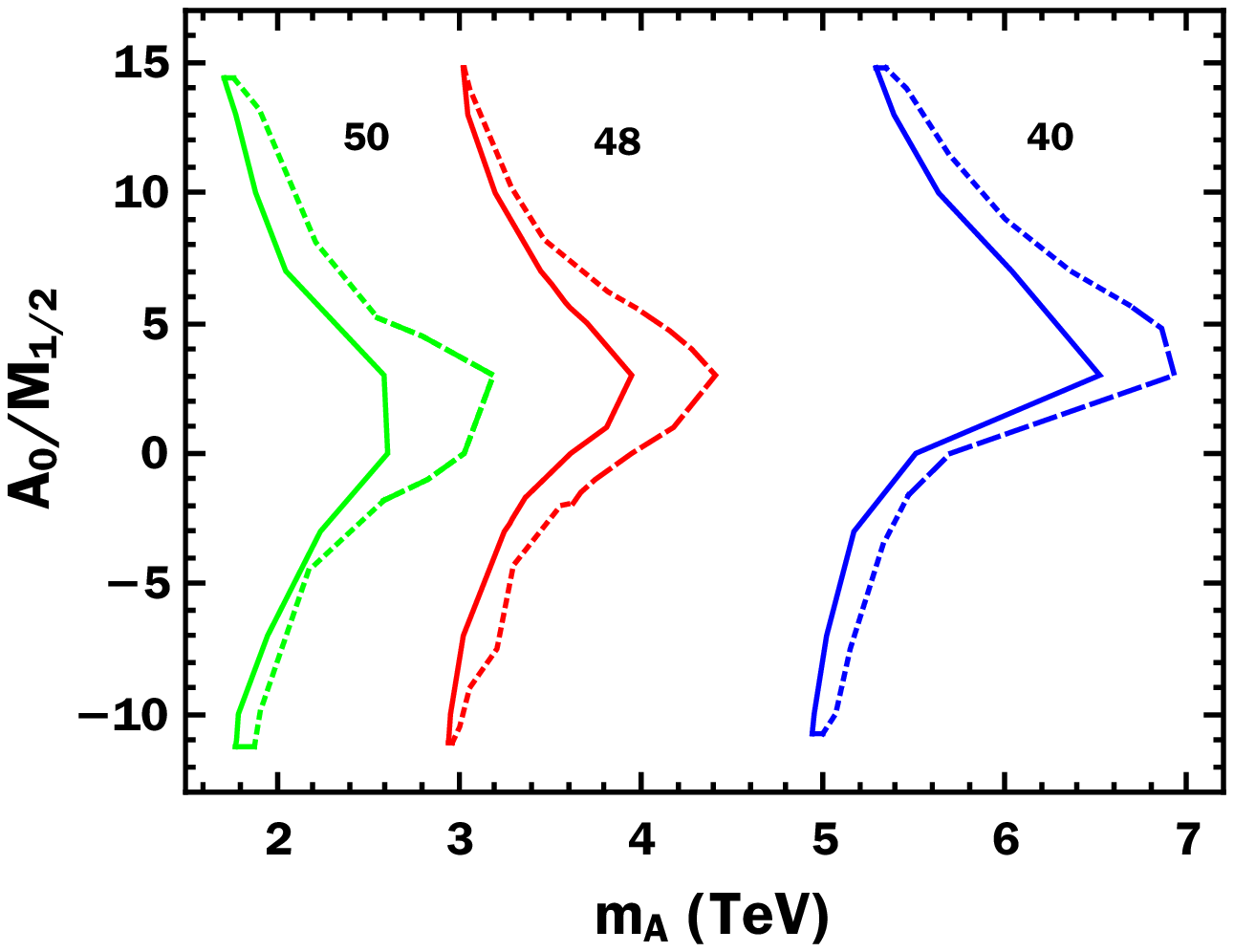,width=9cm}}
\caption{Allowed regions in the $m_A-\amg$ plane with $m_h$
given by \Eref{mhb} and for $\tnb=40 ,48,~\text{and}~50$,
bounded by blue, red, and green lines, respectively. The
solid, dashed, and dotted boundary lines come from the
constraints in Eqs.~(\ref{mchb}), (\ref{omgb}), and the
LUX data, respectively. The restriction in \Eref{mglb}
limits the range of $\amg$, but the corresponding lines
are hardly visible.}
\label{fig:mAA0LUX}
\end{figure}

In order to have a more spherical view of the allowed parameter
space of our model, we present in \Fref{fig:mAA0LUX} the allowed
regions in the $m_A-\amg$ plane for $\tnb=40 ,48,~\text{and}~50$
enclosed by blue, red, and green lines, respectively. As one can
see from \Fref{fig:xisigma} and the allowed ranges of $\mx$
presented in \Sref{reslux}, the mass of the LSP (and $\mg$) is
confined in approximately the same range independently of $\tnb$.
Therefore, we opt to use $m_A$ as variable in the horizontal axis,
so that the allowed regions corresponding to different
$\tan\beta$'s are well distinguishable. We observe that,
although the range of $\mx$ is practically $\tnb$-independent,
$m_A$ increases as $\tnb$ decreases. For each value of $\tan\beta$,
there are three different boundary lines corresponding to different
restrictions. On the solid and dashed line, the bounds on $\mch$
in \Eref{mchb} and on $\omg$ in \Eref{omgb} are saturated, whereas
the restriction from the LUX data on $\xssi$ yields the dotted
boundary line. It is impressive that the LUX data provide such a
strong constraint on the model parameters, which overshadows all
other constraints for approximately $\amg\lesssim-3$ and
$\amg\gtrsim5$. Finally, the boundary lines from the limit on
$\mgl$ in \Eref{mglb}, although hardly visible in this plot,
provide the maximal and minimal $\amg$'s. Note that the
allowed regions are obviously symmetric about $\amg\simeq2.5$.
Also, we find that $\mu$ remains almost constant $\simeq100\pm20
\gev$ on the solid lines from Eq.~(\ref{mchb}), while it reaches
about $1\tev$ when the bound in Eq.~(\ref{omgb}) is saturated.
As regards $\damu$, the acquired values are well below the lower
limit in Eq.~(\ref{damub}). Specifically, in the allowed regions
of \Fref{fig:mAA0LUX}, we obtain $\damu\simeq(0.04-0.27)\times
10^{-10}$. Therefore, \Eref{1damub} is satisfied only at the level
of $2.83$ to $2.86-\sigma$.

The values of the input and some output parameters, the mass
spectra and some low energy observables of the model are listed
in \Tref{tab} for four characteristic points of the allowed
parameter space with $\mh=125.5\gev$. The various masses of the
SUSY particles (gauginos/higgsinos
$\neutralino$, $\neutralino_2$, $\neutralino_3$, $\neutralino_4$,
$\neutralino_1^\pm$, $\neutralino_2^\pm$, $\tilde{g}$, squarks
$\tilde{t}_1$, $\tilde{t}_2$, $\tilde{b}_1$, $\tilde{b}_2$,
$\tilde{u}_L$, $\tilde{u}_R$, $\tilde{d}_L$, $\tilde{d}_R$, and
sleptons $\tilde{\tau}_1$, $\tilde{\tau}_2$, $\tilde{\nu}_\tau$,
$\tilde{e}_L$, $\tilde{e}_R$) and the Higgs particles ($h$, $H$,
$H^\pm$, $A$) are given in $\TeV$ -- note that we consider the
first two generations of squarks and sleptons as degenerate. We
chose the values of the input parameters so as to assure that
$\omg$ is not far from its central value in \Eref{eq:omegalimit}.
The relatively low $\dch$'s and $\dnt$'s obtained and the sizable
higgsino fraction $|N_{1,3}|^2+|N_{1,4}|^2$  assist us to achieve
this. Note that, in all the cases, the lightest neutralino mass
turns out to be close to the corresponding $\mu$. We also include
an estimate for the EWSB fine-tuning parameter $\dew$ -- see
\Sref{nat}.

We observe that the low energy $B$-physics observables satisfy the
relevant constraints in \Sref{pbph}, whereas the $\damu$'s are far
below the lower limit in \Eref{damub}. Also, the extracted
$\ssi$'s can be probed by the forthcoming experiments
\cite{xenon1t, supercdms}. For the spin-dependent $\xx-p$
scattering cross section $\ssd$, we adopt the central values for
the hadronic inputs in \cref{karagiannakis2011}. As can be easily
deduced from the displayed values, $\ssd$ in our model lies well
below the sensitivity of IceCube \cite{icecube} (assuming
$\xx-\xx$ annihilation into $W^+W^-$) and the expected limit from
the large DMTPC detector \cite{brown}. Therefore, the LSPs
predicted by our model can be detectable only by the future
experiments which will release data on $\ssi$.

Comparing the results of \Tref{tab2} with the corresponding ones in
the $\xstau$ coannihilation region -- see Table~II of
Refs.~\cite{karagiannakis2012,karagiannakis2013} --, we may
appreciate the different features of the HB/FP solutions presented
here. First of all, $m_0$ acquires considerably larger values here,
while  $\mu$ remains quite small. The Higgs bosons $H$, $H^\pm$, and
$A$ acquire larger masses and the whole sparticle spectrum, with the
exception of the neutralinos and charginos, becomes heavier. The
gluino mass is heavier for lower $\tan\beta$'s, in contrast with
what happens in the $\xstau$ coannihilation region, where we observe
the opposite behavior. Here, the ratio $h_b/h_\tau$ is smaller,
$h_t/h_b$ is closer to 2, while $h_t/h_\tau$ is even closer to unity.
The restrictions on the low energy observables are all well satisfied,
except the one on $\damu$, which becomes even smaller than in the
$\xstau$ coannihilation region. Note that the latter region is
tightly constrained by $\bsmm$, which is well suppressed in the HB/FP
region. As regards $\omg$, this becomes compatible with \Eref{omgb}
in the present case thanks to $\xx/\xxb-\cha$ coannihilations, which
are activated because of the low $\dch$'s and $\dnt$'s. Note that,
in the $\xstau$ coannihilation region, the relevant mass splittings
are $\dstau=(m_{\tilde\tau_1}-m_{\rm LSP})/m_{\rm LSP}$ and
$\Delta_H=(m_H-2m_{\rm LSP})/2m_{\rm LSP}$. Finally, here we
obtain a large higgsino fraction of the LSP, which confines
$\mx$ close to $\mu$ and as a bonus ensures $\ssi$'s accessible to
the  forthcoming experiments \cite{xenon1t,supercdms}. On the
contrary, within the $\xstau$ coannihilation region, $\xx$ is an
almost pure bino, $\mx\simeq\mg/2$, and $\ssi$ is well below the
sensitivity of any planned experiment.

\renewcommand{\arraystretch}{1.1}
\begin{table*}[t!]
\caption{Input/output parameters, sparticle and Higgs masses, and
low energy observables in four cases in the HB/FP.} \label{tab}
\begin{ruledtabular}
\begin{tabular}{c@{}c@{}c@{}c@{}c}
\multicolumn{5}{c}{Input parameters}\\\hline
%-------------------------------------------------------------------------
$\tan\beta$ & $40$ &$45$ & $48$ & $50$\\
$\amg$ &$2$&$0$ &$-1.5$&$-3$ \\
$M_{1/2}/\TeV$ &$3.763$& $2.945$ &$2.161$&$1612.4$ \\
$m_0/\TeV$&$9.603$ &$8.821$ &$9.231$&$9.561$ \\ \hline
%--------------------------------------------------------------------------
\multicolumn{5}{c}{Output parameters}\\\hline
$h_t/h_\tau(M_{\rm GUT})$&$1.474$ & $1.237$ & $1.107$ & $1.027$ \\
$h_b/h_\tau(M_{\rm GUT})$&$0.756$ & $0.758$ & $0.763$ & $0.774$ \\
$h_t/h_b(M_{\rm GUT})$&$1.949$ & $1.631$ & $1.45$ & $1.326$ \\\hline
$\mu/\TeV$ &$1.018$& $1.01$ & $0.928$ & $0.736$ \\
$\dch$&$0.002$ & $0.004$ & $0.014$ & $0.038$ \\
$\dnt$&$0.004$ & $0.006$ & $0.018$ & $0.045$ \\
$\dew$&$244.6$&$237.9$  &$216.2$ &$130.9$\\ \hline\\[-2.5mm]
$|N_{1,3}|^2+|N_{1,4}|^2$&$0.997$&$0.992$  &$0.836$
&$0.54$\\\hline
\multicolumn{5}{c}{Sparticle and Higgs boson masses
in$\tev$}\\\hline
$\tilde\chi, \tilde\chi_2$&$1.043, 1.047$& $1.026, 1.032$ &$0.935, 0.952$ &$0.723, 0.756$\\
$\tilde{\chi}_{3}, \tilde{\chi}_{4}$&$1.792, 3.381$ &$1.397, 2.651$ &$1.034, 1.959$ &$0.788,
1.473$\\
$\tilde{\chi}_{1}^{\pm}, \tilde{\chi}_{2}^{\pm}$&$1.046, 3.381$ &$1.030, 2.651$ &$0.949, 1.959$
&$0.751, 1.473$\\
$\tilde{g}$&$8.147$&$6.512$ &$4.951$ &$3.799$\\ \hline\\[-2.5mm]
%------------------------------------------------------------------------
%
$\tilde{t}_1, \tilde{t}_2$&$8.113, 9.516$ &$6.900, 7.984$ &$6.309, 7.270$ &$6.049, 6.900$ \\
$\tilde{b}_1, \tilde{b}_2$&$9.515, 10.23$ &$7.987, 8.560$ &$7.267, 7.887$ &$6.897, 7.512$\\
$\tilde{u}_{L}, \tilde{u}_{R}$&$11.777, 11.553$&$10.357, 10.197$ &$10.100, 10.004$ &$10.082,
10.022$\\
$\tilde{d}_{L}, \tilde{d}_{R}$&$11.777, 11.524$&$10.357, 10.176$ &$10.100, 9.992$ &$10.082,
10.014$\\\hline
%--------------------------------------------------------------------------
$\tilde\tau_1, \tilde\tau_2$&$7.927, 9.104$&$6.921, 8.136$ &$6.749, 8.202$ &$6.620, 8.292$\\
$\tilde\nu_\tau$&$9.103$ &$8.135$ &$8.201$ &$8.291$\\
$\tilde{e}_L, \tilde{e}_R$&$9.936, 9.718$&$9.050, 8.900$ &$9.359, 9.276$ &$9.640, 9.589$\\
$\tilde{\nu}_{e}$&$9.935$ &$9.049$ &$9.359$ &$9.639$\\\hline
%--------------------------------------------------------------------------
%
$h, H$&$0.1255, 6.56$&$0.1255, 4.820$ &$0.1255, 3.67$ &$0.1255, 2.369$\\
$H^{\pm}, A$&$6.56, 6.56$&$4.820, 4.820$  &$3.671, 3.67$ &$2.370, 2.371$ \\\hline
\multicolumn{5}{c}{Low energy observables}\\\hline\\[-2.5mm]
%--------------------------------------------------------------------------
%
$10^4\bsg$&$3.31$ &$3.30$  &$3.30$ &$3.33$\\
$10^9\bsmm$ &$3.04$ &$3.03$ &$3.02$ &$2.97$\\
$\btn$ &$0.998$ &$0.995$  &$0.991$ &$0.976$\\
$10^{10}\damu$ &$0.135$ &$0.2$&$0.227$ &$0.237$\\\hline\\[-2.5mm]
$\omg$ &$0.11$&$0.11$&$0.11$&$0.11$\\[0.8mm]
$\ssi / 10^{-9} \mathrm{pb}$&$0.28$ &$0.81$&$7.75$ &$13.29$\\[1mm]
$\ssd / 10^{-7} \mathrm{pb}$&$ 2.55$ &$ 7.87$&$77.08$
&$227.8$\\
\end{tabular}
\end{ruledtabular}
\label{tab2}
\end{table*}

\begin{figure*}[ht!]
\centerline{\epsfig{file=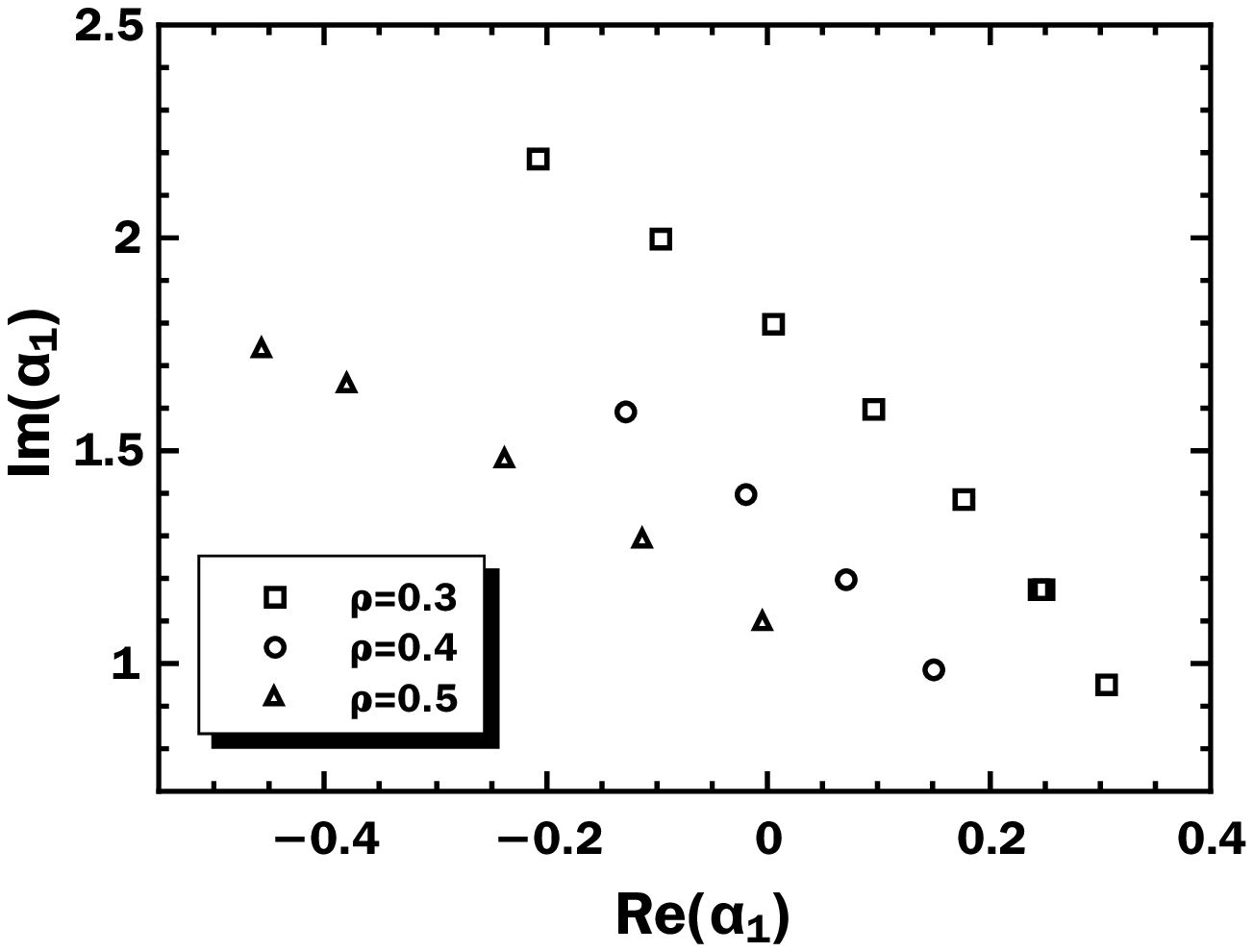,width=9cm}
\epsfig{file=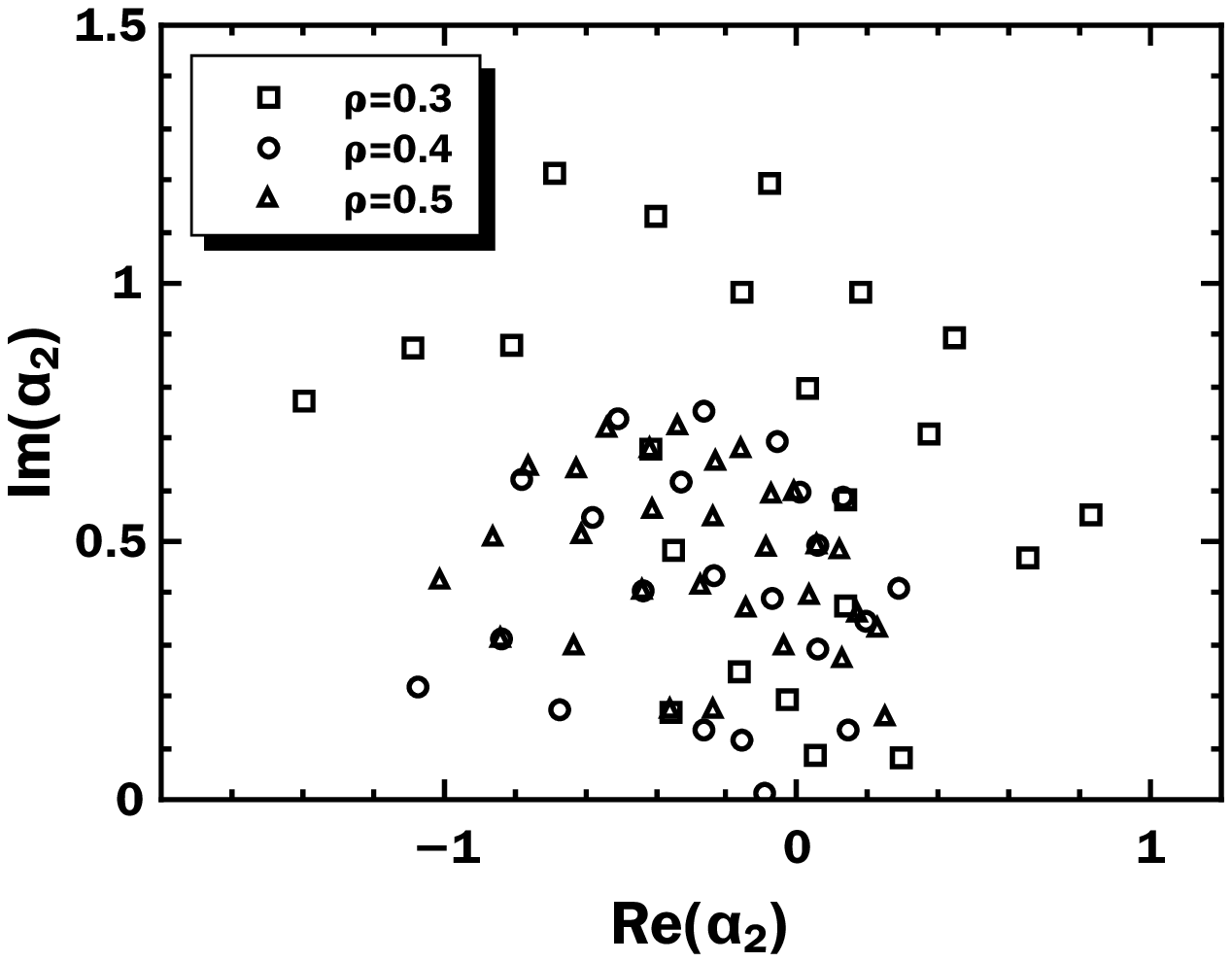,width=9cm}} \caption{\footnotesize The
complex parameters $\alpha_1$ and $\alpha_2$ for various real and
positive values of $\rho$ indicated on the graphs for the case in
the third column of \Tref{tab} with $\tnb=48$.}\label{fig:ayukawa}
\end{figure*}

\section{Deviation from Yukawa Unification} \label{yuk}

The ranges of the ratios of the asymptotic third generation
Yukawa coupling constants in the allowed parameter space of
the model in the range $40\leq\tnb\leq50$, which we consider
here, are the following:
\beq 1\lesssim
\frac{h_t}{h_\tau}\lesssim1.5,~0.75\lesssim
\frac{h_b}{h_\tau}\lesssim0.79,~\text{and}~
1.2\lesssim\frac{h_t}{h_b}\lesssim2.
\label{specratio}\eeq
It is easy to see that, although YU is violated, these ratios
turn out to be quite close to unity. However, for $\tan\beta<40$,
the ratios $h_t/h_\tau$ and $h_t/h_b$ become large and the term
`Yukawa quasi-unification conditions' cannot be justified for
the constraints given in \Eref{yquc}. So, we restrict ourselves
to $\tnb\geq40$.

We will now give a specific example of how such values for the
ratios of the Yukawa coupling constants as the ones in
Eq.~(\ref{specratio}) can be achieved in a natural way. We
consider the third case in \Tref{tab} with $\tan\beta=48$, where
$h_t/h_\tau=1.107$ and $h_b/h_\tau=0.763$, and solve \Eref{yquc}
with respect to the complex parameters $\alpha_1$, $\alpha_2$, and
the real and positive parameter $\rho$. Specifically, we first
find pairs of values for $\rho$ and $\alpha_1$ which satisfy the
YQU condition for $h_b/h_\tau$ and, then, for each one of these
pairs, we find $\alpha_2$'s which satisfy the equation for
$h_t/h_\tau$. Note that, while $h_b/h_\tau$ depends only on the
value of the product $\rho\alpha_1$, the ratio $h_t/h_\tau$
depends on each of the parameters $\alpha_1$, $\alpha_2$, and
$\rho$ separately. This is the reason why the solutions in the
$\alpha_1$ complex plane lie on a set of similar curves
corresponding to different values of $\rho$, whereas the values of
$\alpha_2$ do not follow any specific pattern in the corresponding
complex plane -- see below. We found that solutions exist only for
low values of $\rho$ (up to about 0.6) and also for large values
of this parameter ($\rho\gtrsim2.4$).

We present several of these solutions in the $\alpha_1$ and
$\alpha_2$ complex planes in \Fref{fig:ayukawa}, but only for the
lower values of $\rho$, which are considered more natural. In the
left panel of this figure, we see that, for any given value of
$\rho$, the data clearly lie on a specific curve and that the
curves for different values of $\rho$ are similar to each other.
On the other hand, in the right panel, the data are more
complicatedly distributed. Note that, every pair of values of
$\alpha_1$ and $\rho$ corresponds to more than one value of
$\alpha_2$. From Fig.~\ref{fig:ayukawa}, we can deduce that, for
the specific example considered and for small values of $\rho$,
the ranges of the values of all the parameters are the following:
\beqs\begin{align}
0.3\lesssim ~& \rho~\lesssim0.5,\\
-0.45\lesssim\Re(\alpha_1)\lesssim0.31,~&0.95
\lesssim\Im(\alpha_1)\lesssim2.19,\\
-1.39\lesssim\Re(\alpha_2)\lesssim0.83,~&0.01
\lesssim\Im(\alpha_2)\lesssim1.21.
\end{align}\eeqs
Note also that, for the larger values of $\rho$, we get smaller
values for $\alpha_1$ and $\alpha_2$. So, we see that the ratios
of the Yukawa coupling constants in this example can be easily
obtained by natural choices of $\rho$, $\alpha_1$, and $\alpha_2$.

We find that, for all the possible values of the ratios of the
third generation Yukawa coupling constants encountered in our
investigation, the picture is quite similar. So, we conclude
that these ratios can be readily obtained by a multitude of
natural choices of the parameters $\rho$, $\alpha_1$, and
$\alpha_2$ everywhere in the allowed parameter space of the
model which we considered here.

\section{Naturalness of the EWSB}\label{nat}

One of the most important reasons for introducing SUSY was that
it could provide a solution to the hierarchy problem. However,
the fact that, in our model, $m_0$ turns out to lie in the
multi-$\TeV$ range generates doubts about the naturalness of
the radiative EWSB, since it leads to the reappearance of a
small fine-tuning problem. This is the so-called \textit{little
hierarchy problem}, which is still a much debated issue.

To quantify somehow the naturalness of our model with respect
to this issue, we focus on the EWSB condition relating $M_Z$
to $\mu$ and $\tan\beta$. This condition is obtained by
minimizing the tree-level renormalization-group improved
scalar potential for $H_1$ and $H_2$ and reads as follows
\cite{barger1994}
\begin{equation}
\label{min}
\frac{1}{2}M_Z^2\simeq\frac{m_{H_1}^2-m_{H_2}^2\tan^2\beta}
{\tan^2\beta-1}-\mu^2,
\end{equation}
where $m_{H_1}$ is the soft SUSY breaking mass of $H_1$ and
we neglect possible loop corrections, which are anyway
minimized thanks to the choice of the optimal scale in
\Eref{opt}. In the case where the value of the fraction in the
right-hand side of Eq.~(\ref{min}) is quite large, a large
value of $\mu^2$ is also needed. However, this requires large
values of $m_0$ and $M_{1/2}$ and heavy sparticle spectrum and,
therefore, introduces a certain amount of fine-tuning. To
measure this tuning, we introduce the EWSB fine-tuning
parameter
\begin{equation}
  \dew\equiv\mathrm{max}\left(\frac{|C_i|}{M_Z^2/2}\right),
\end{equation}
where the $C_i$'s ($i=\mu,H_1,H_2$) are -- see
\Eref{min} -- :
\beq \lf C_{\mu}, C_{H_1}, C_{H_2}\rg= \lf-\mu^2,
\frac{m_{H_1}^2}{\tan^2\beta-1},
-\frac{m_{H_2}^2\tan^2\beta}{\tan^2\beta-1}\rg\cdot \eeq
In most of the parameter space explored, $\dew$ is dominated
by the term $C_\mu$.

\begin{figure*}[t!]
\centerline{\epsfig{file=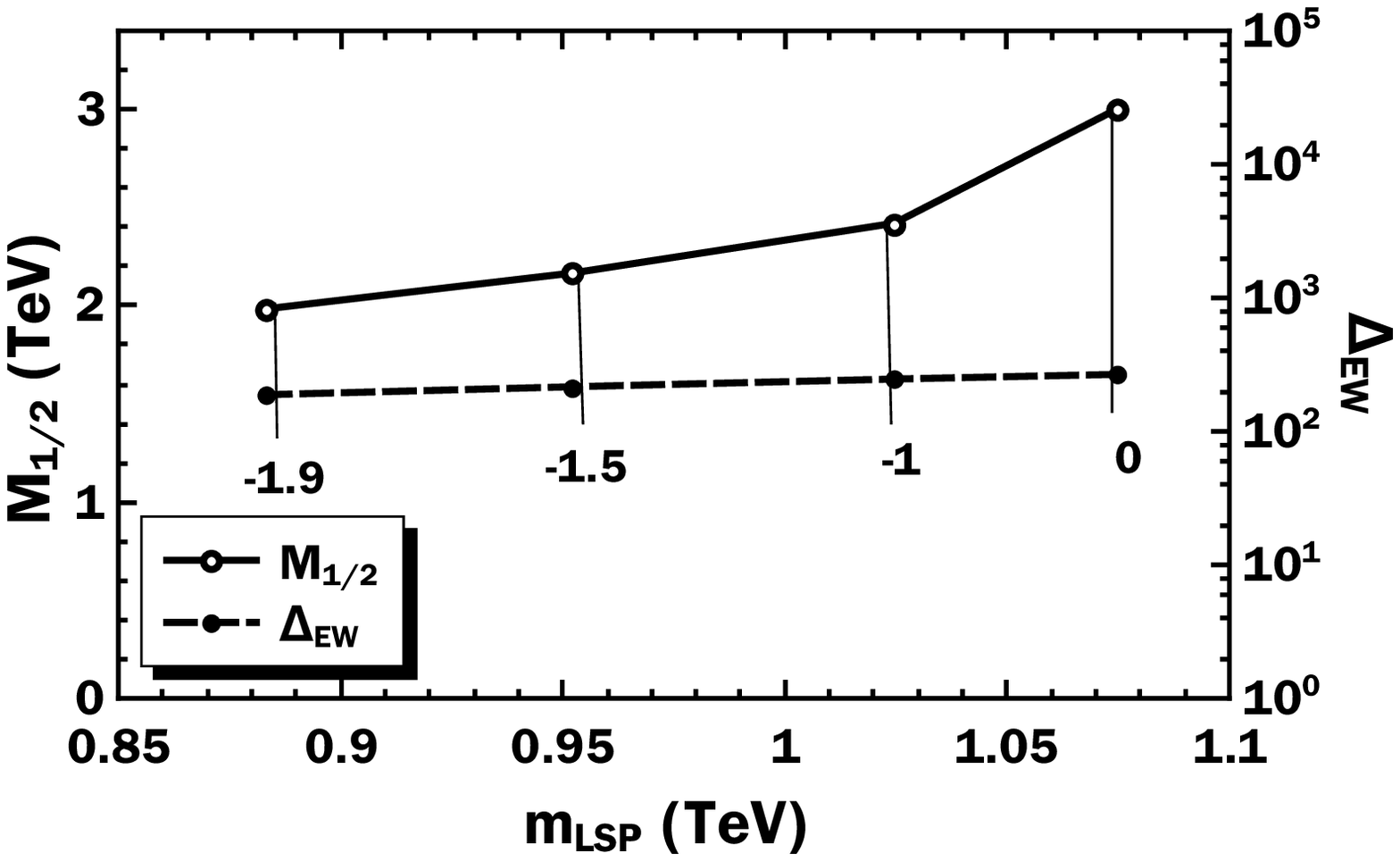,width=9cm}
\epsfig{file=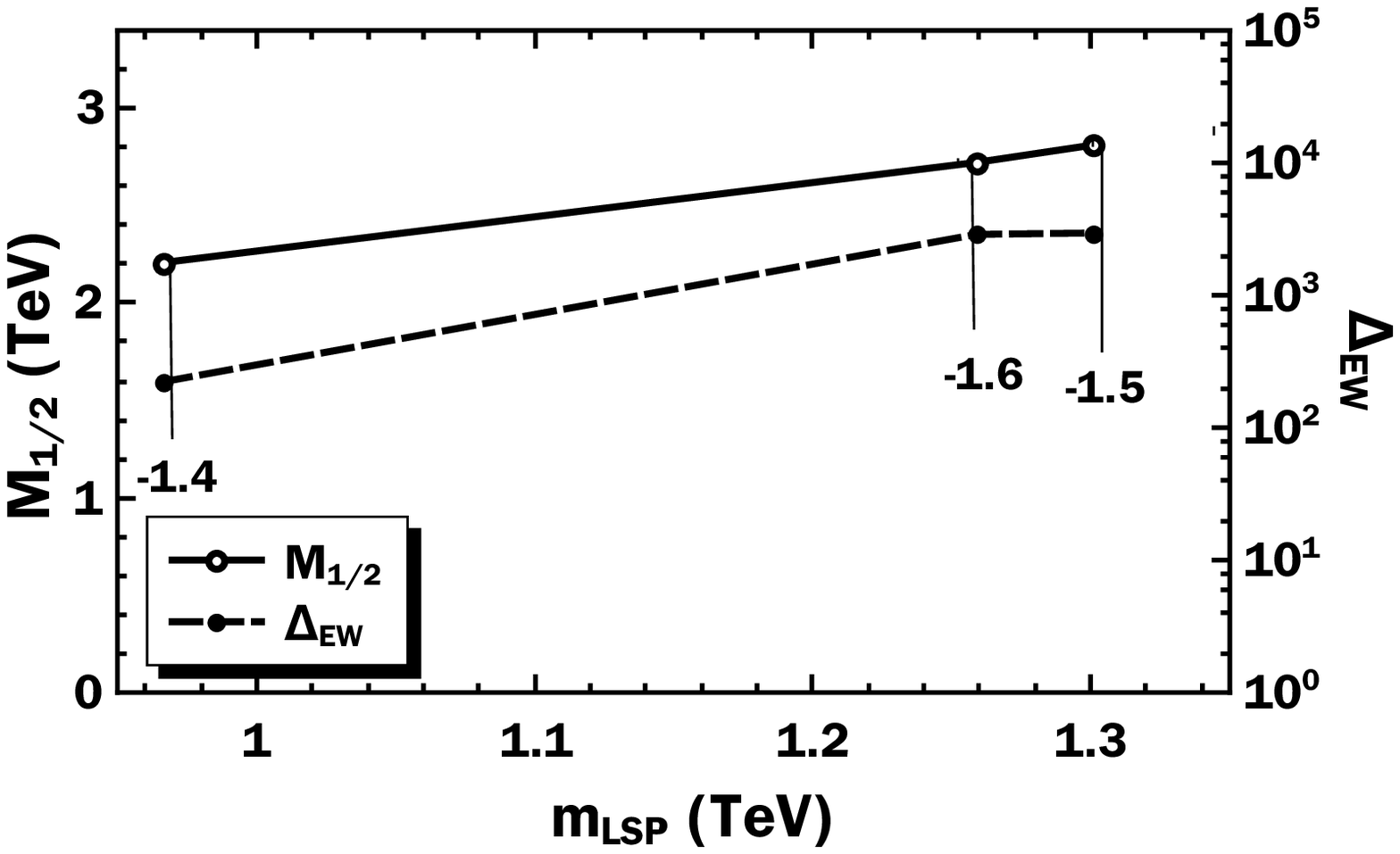,width=9cm}}
\caption{\footnotesize $M_{1/2}$ and $\dew$ as functions of $\mx$
for $\omg=0.125$, $\mh$ given by \Eref{mhb} and various $\amg$'s
indicated on the curves for the HB/FP (left panel) and the
$\xstau$ coannihilation (right panel) region of the model.}
\label{fig:finetuning}
\end{figure*}

Focusing on the values of the parameters which ensure
$\omg=0.125$, we present, in the left panel of
\Fref{fig:finetuning}, $\mg$ (solid line) and $\dew$ (dashed line)
as functions of $\mx$ for $\tnb=48$, $\mh=125.5~\GeV$, and
negative values of $\amg$ indicated in the graph -- cf. the left
panel in \Fref{fig:xisigma} for $\tnb=48$. We clearly see that the
required EWSB fine-tuning parameter $\dew$ is almost constant and
$\sim 200$. This result is also valid for the other two values of
$\tnb$ in \Fref{fig:xisigma}. Indeed, for the largest value of
$|\amg|$ on the dashed line in each of the six panels of
\Fref{fig:xisigma}, the derived $\dew$ is as follows:

\begin{itemize}
\item For $\tan\beta=40$ and $\amg=-1.6$ [$\amg=5.7$], we
find $\dew=191.311$ [$\dew=187.379$].

\item For $\tan\beta=48$ and $\amg=-1.9$ [$\amg=5.6$], we
find $\dew=188.064$ [$\dew=187.724$].

\item For $\tan\beta=50$ and $\amg=-1.8$ [$\amg=5.2$], we
find $\dew=204.702$ [$\dew=201.228$].

\end{itemize}

The value of $\dew$ can be even smaller for lower $\mx$'s and
$\omg$'s. However, we believe that the values presented above
are more interesting, especially if we wish to compare the
$\dew$'s found in the HB/FP region with the ones in the
$\xstau$ coannihilation region analyzed in
\cref{karagiannakis2012}. To this end, we plot also, in
the right panel of \Fref{fig:finetuning}, $\mg$ (solid line)
and $\dew$ (dashed line) as functions of $\mx$ in the $\xstau$
coannihilation region of the same model for $\tnb=48$,
$\mh=125.5~\GeV$, and $\amg<0$ -- see Fig.~3 of
\cref{karagiannakis2012}. It is evident that, in this
coannihilation region, the required values of $\mx$ are
somewhat larger and the values of $\dew$ can become about a
factor of ten larger ($\dew>1000$). We can, thus, easily
appreciate the amelioration regarding the EWSB fine-tuning
that we achieve working in the HB/FP region.

\section{Conclusions}\label{con}

We investigated the compatibility of the generalized asymptotic
Yukawa coupling constant quasi-unification, which yields
acceptable masses for the fermions of the third family, with
the HB/FP
region of the CMSSM for $\mu>0$ and $40\leq\tnb\leq50$. We
imposed phenomenological constraints originating from the mass
of the lightest neutral CP-even Higgs boson, the lower bounds
on the masses of the sparticles, and $B$-physics. We also
considered cosmological constraints coming from the CDM relic
abundance in the universe and the LUX data on the
spin-independent neutralino-proton elastic scattering cross
section. Fixing $\mh$ to its central value favored by the LHC,
we found a relatively wide allowed range of parameters with
$-11\lesssim\amg\lesssim15$ and $0.09\lesssim\mx/\TeV\lesssim1.1$.
The restriction on the deviation of the measured value of the
muon anomalous magnetic moment from its SM prediction, however,
is only satisfied at the level of about $2.8-\sigma$ in this
parameter space allowed by all the other requirements.

The LSP, which is a bino-higgsino admixture, has an acceptable
relic abundance thanks to coannihilations between the LSP and the
next-to-lightest neutralino with the lightest chargino. The LSP is
also possibly detectable in the planned CDM direct search
experiments which look for the spin-independent elastic cross
section between neutralino and proton. The required deviation from
YU can be easily attributed to a multitude of natural values of
the relevant parameters within a Pati-Salam SUSY GUT model and the
EWSB fine-tuning $\dew^{-1}$ turns out to be of the order of
$5\times10^{-3}$. It is worth mentioning that the same model has
been tested successfully in the $\xstau$ coannihilation region
\cite{karagiannakis2012}. However, the allowed parametric space
turned out to be much more restricted there, the detectability of
the LSP quite difficult, and the EWSB fine-tuning worse.

\paragraph*{\textbf{Acknowledgements} } We would like to thank B.E.
Allanach and D.~George in DAMTP, Cambridge as well as
N.~Mavromatos in King's College, London for useful discussions.
This research was supported from the MEC and FEDER (EC) grants
FPA2011-23596 and the Generalitat Valenciana under grant
PROMETEOII/2013/017.

\def\plb#1#2#3{{Phys. Lett. B }{\bf #1}, #3 (#2)}
\def\prl#1#2#3{{Phys. Rev. Lett.}
{\bf #1},~#3~(#2)}
\def\prd#1#2#3{{Phys. Rev. D }{\bf #1}, #3 (#2)}
\def\npb#1#2#3{{Nucl. Phys. }{\bf B#1}, #3 (#2)}
\def\jhep#1#2#3{{J. High Energy Phys.}
{\bf #1}, #3 (#2)}

\end{document}